\documentclass[a4paper,11pt]{article}
\pdfoutput=1
\usepackage{jinstpub}
\usepackage{dcolumn}
\usepackage{hyperref}
\usepackage{multirow}
\usepackage{color}
\usepackage{url}
\usepackage{xcolor}
\usepackage{tabularx}
\usepackage{array,booktabs}
\usepackage{caption}
\usepackage{subcaption}
\graphicspath{{figs/}}  
\usepackage{adjustbox}
\usepackage{placeins}
\usepackage{lineno}

\definecolor{BlueViolet}{rgb}{0.2, 0.00, 0.7}
\definecolor{Blue}{rgb}{0.15, 0.00, 0.9}
\definecolor{mj-oj}{RGB}{209, 94, 6}

\title{Photomultiplier Requirements and Pre-Calibration for the SABRE South Liquid Scintillator Veto} 

 \affiliation[1]{School of Physics, The University of Melbourne, Parkville, VIC 3010, Australia}
\affiliation[2]{Department of Nuclear Physics and Accelerator Applications, The Australian National University, Canberra, ACT 2601, Australia}
\affiliation[3]{Department of Physics, The University of Adelaide, Adelaide, SA 5005, Australia}
 \affiliation[4]{ARC Centre of Excellence for Dark Matter Particle Physics, Australia}
\affiliation[5]{Department of Physics, The University of Toronto, ON M5R 2M8, Canada}
 \affiliation[6]{School of Physics, The University of Sydney, NSW 2006 Camperdown, Sydney, Australia}
 \affiliation[7]{
International Center for Quantum-field Measurement Systems for Studies of the Universe and Particles
(QUP), High-Energy Accelerator Research Organization (KEK), Oho, Tsukuba, Ibaraki 305-0801, Japan}
 \affiliation[8]{INFN Sezione di Milano, via Celoria 16, 20133 Milano, Italy}

\author[1,4,a]{L.~J.~Milligan,}
\author[1,4,b]{P.~Urquijo,}
\author[1,4]{E.~Barberio,}
\author[2,4]{V.~U.~Bashu,}
\author[2,4]{L.~J.~Bignell,}
\author[3,4,8]{I.~Bolognino,}
\author[1,4]{S.S.~Chhun,}
\author[2,4]{F.~Dastgiri,}
\author[6,4]{T.~Fruth,}
\author[1,4]{G.~Fu,}
\author[3,4]{G.~C.~Hill,}
\author[1,4]{Y.~Hua,}
\author[1,4]{R.~S.~James,}
\author[3,4]{K.~Janssens,}
\author[6,4]{S.~Kapoor,}
\author[2,4]{G.~J.~Lane,}
\author[3,4]{K.~T.~Leaver,}
\author[3,4]{P.~McGee,}
\author[2,4]{L.~J.~McKie,}
\author[1,4]{J.~McKenzie,}
\author[2,4,5]{P.~C.~McNamara,}
\author[1,4]{W.~J.~D.~Melbourne,}
\author[1,4]{M.~Mews,}
\author[1]{W.~H.~Ng,}
\author[1,4]{K.~J.~Rule,}
\author[2,4]{Z.~Slavkovsk\'{a},}
\author[1,4]{O.~Stanley,}
\author[2,4]{A.~E.~Stuchbery,}
\author[7]{B.~Suerfu,}
\author[1,4]{G.~N.~Taylor,}
\author[2]{D.~Tempra,}
\author[2]{T.~Tunningley,}
\author[3,4]{A.~G.~Williams,}
\author[2,4]{Y.~Y.~Zhong,}
\author[4,5]{M.~J.~Zurowski}

\collaboration{The SABRE South Collaboration}

\collaboration{SABRE South Collaboration}
\emailAdd{sabre-contact@lists.unimelb.edu.au, purquijo@unimelb.edu.au}
\abstract{
    We present a study of the  oil-proof base Hamamatsu R5912 photomultiplier tubes that will be used in the SABRE South linear-alkylbenzene liquid scintillator veto. SABRE South is a dark matter direct detection experiment at the Stawell Underground Physics Laboratory, aiming to test the DAMA/LIBRA dark matter annual modulation signal. We discuss the requirements of the liquid scintillator system and its photomultipliers, outline the methods and analysis used for the characterisation measurements, and results from initial tests. We discuss the impact of these measurements on the performance of the active veto system and explore analysis methods to allow for low threshold operation. Finally, we include results from a small scale liquid scintillator detector prototype used to assess the future performance of pulse shape discrimination in the liquid scintillator veto, and how well accommodated it is by the R5912 PMTs. 
}

\begin{document}

\maketitle
\flushbottom

\section{Introduction}

This paper presents the characterisation of the oil-proof base Hamamatsu Photonics R5912 photomultiplier tubes (PMTs) for the veto detector of the SABRE South dark matter experiment, including testing in a liquid scintillator prototype detector and the development of analysis methods for background rejection and particle identification.

The SABRE experiment aims to provide a test of the long-standing dark matter signal claim by the DAMA experiment \cite{Bernabei:2018epj}. SABRE consists of two detectors --- SABRE North at Laboratori Nazionali del Gran Sasso (LNGS), Italy and SABRE South at the Stawell Underground Physics Laboratory (SUPL)~\cite{TDRSUmm}, Australia \cite{SABRE_POP} (the first underground laboratory in the Southern Hemisphere). The use of detectors located in two separate hemispheres will allow for the exclusion of seasonal systematic effects.  At the heart of SABRE South is an array of high-purity NaI(Tl) crystals coupled to pairs of 76~mm diameter R11065 PMTs (Hamamatsu). These crystal detector modules are mounted in the centre of a stainless steel vessel which contains 12~kL of linear-alkylbenzene liquid scintillator and is instrumented with a minimum of eighteen 204~mm diameter R5912 PMTs from Hamamatsu. This is the largest liquid scintillator veto amongst similar NaI(Tl) experiments. A total of 32 PMTs could be installed in the veto, given we have recently acquired 16 R5912 PMTs from the decommissioned Daya Bay outer detector~\cite{DayaBay}. The liquid scintillator veto, capable of tagging external and intrinsic background processes, is one of the chief improvements of the SABRE South detector over other NaI(Tl) dark matter searches~\cite{COSINE2024, Coarasa:2024xec}. The main role of this liquid scintillator veto is to tag radioactive decays within the crystals, but will also act as passive shielding for external background. In addition, a top layer of eight plastic scintillator panels located above the vessel serves as a muon veto. 

The linear-alkylbenzene is nominally doped with 3.5~g/L of the fluorophore 2,5-Diphenyloxazole (PPO) and 15~mg/L of Bis-MSB to mitigate the self-absorption properties of LAB and to maximise the total light yield \cite{Anderson_2021}. The LAB emission spectrum peaks at 280~nm \cite{lab-lambda}, through subsequent absorption and emission by first PPO and then Bis-MSB, the emission wavelength is shifted to a spectrum centred around 422~nm, which is better matched to the maximum PMT quantum efficiency at 390~nm.
Light collection efficiency is improved by lining the internal vessel walls with sheets of reflective polyester (PET) film, or ``Lumirror''. The locations of the PMTs in the veto vessel, and their incident angles were chosen so that they point toward the centres of the crystal detectors to maximise detection efficiency. Two calibration systems are installed in the sub-detector: a radioactive source based calibration system and an optical calibration system that uses a picosecond pulsed laser source. The radioactive calibration system is used for energy scale calibration while the optical calibration system serves to monitor the performance of the PMTs over time.

 Simulation studies have shown that effective background reduction for processes that occur in the energy region-of-interest (e.g. intrinsic $^{40}$K decays) is achieved with a veto detector energy threshold of a few hundred keV~\cite{sabre_background}, and it may be possible to push the threshold as low as 50~keV \cite{SABRE_MC,sabre_background,TDRSUmm}. However, in the case of 32 PMTs, a threshold as low as 20 keV becomes increasingly achievable. This lower threshold would open up the possibility of using the liquid scintillator system to detect unusual dark matter signals such as those described in Refs.~\cite{iBDM,mimps}, as well as supernova neutrinos and participation in a supernova early warning system (SNEWS)~\cite{SNEWS:2020tbu}. To achieve a low energy threshold throughout the volume of the detector the photon detection efficiency must be maximised given a light yield of $\mathcal{O}(10)$~PE/keV (photoelectrons/keV). Simultaneously, dead-time due to external background must be minimised, and the signal-to-noise ratio must be well understood.

The PMTs used in the liquid scintillator veto detector must be calibrated prior to installation. Each PMT's properties have important implications for how efficiently background can be vetoed and how accurately they can be characterised. The following characterisation measurements have been performed and will be described in detail in this paper: dark rate as a function of temperature, single photon-electron response in terms of charge and timing, quantum efficiency, and the dynamic range of the PMT. In addition to this, spontaneous light emission in the oil-proof base of the PMT is studied. This is a known noise effect reported by other experiments using similar potted bases~\cite{DoubleChooz:2016ibm}.

The liquid scintillator veto sub-detector can also be used to characterise background contributions. A key aspect of background reconstruction is particle identification (PID) based on digitised waveform information. To develop and understand the viability of new PID methods that could be applied to the full scale detector, a small prototype using the same liquid scintillator and PMT model is studied. Machine learning methods for gamma-ray/neutron discrimination are presented. This technique shows promise and could be adapted to the final detector to great effect.

\section{Veto Sub-detector Requirements and Simulation}

\subsection{Simulation and Geometry}
\begin{figure}[htb]
   \centering
   \includegraphics[width=0.5\textwidth]{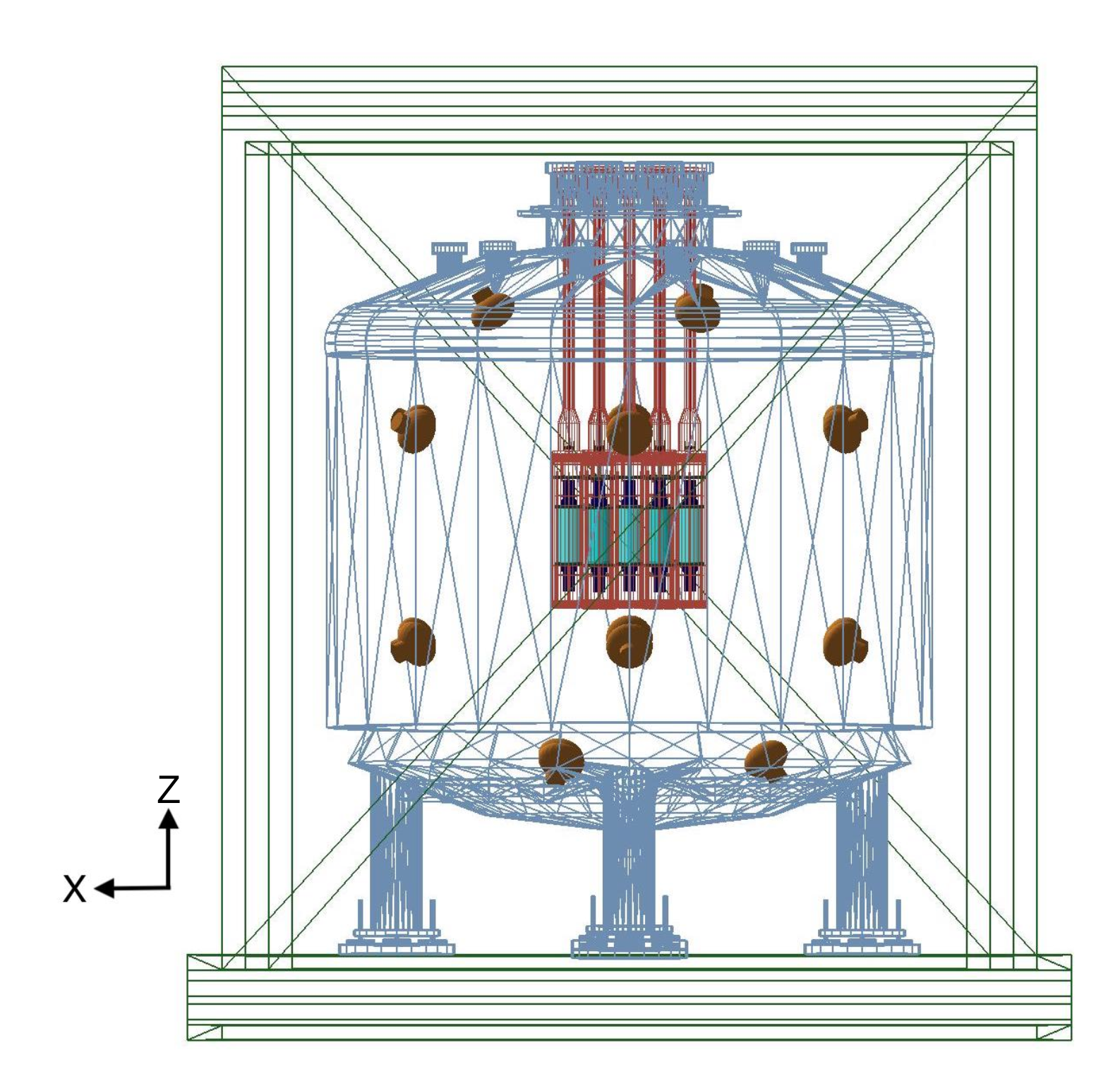}
   \caption{Simulated version of the SABRE South detector. The coordinates labelled correspond to the internal coordinate system of the Geant4 setup (described in text).}
   \label{fig:southgeo}
\end{figure}

Simulations based on Geant4~\cite{GEANT4} are used to define the requirements of the SABRE South PMTs and the pre-calibration process. A render of the geometry is shown in Fig.~\ref{fig:southgeo}. The coordinate system is defined with the origin being the centre of the crystal in the middle of the array, and the Z axis is the vertical direction. The R5912 PMTs are modelled as a 2~mm thick ellipsoid window made of borosilicate glass, with a 10~$\mu$m ceramic inner surface to serve as a photocathode.

To produce simulations of optical photons within the veto sub-detector, we need to assume the light yield (LY) of the liquid scintillator. The liquid scintillator is given a LY of 12 PE/keV (on par with measurements of other LAB based scintllators~\cite{JUNOPhysics, LABChar}), with a final emission probability equal to that of the Bis-MSB emission. The Lumirror reflectivity spectrum used is flat at $\sim$90\% within the range of Bis-MSB emission.

The following pieces of information are used to simulate the PE statistics and photon detection probabilities.
\begin{enumerate}
   \item The number of optical photons generated by a deposition of energy $E$, which is probabilistic but governed by the light yield, $\rm LY$. Thus, a Poisson distribution is used such that the probability that $n$ photons are generated is
       \begin{equation}
           {\rm Pois}(n;{\rm LY}\times E) = \frac{({\rm LY}\times E)^n \exp[-{\rm LY}\times E]}{n!}.
       \end{equation}
   \item The probability that an optical photon generated at position $(x,y,z)$ reaches a given PMT$_i$. This is strongly dependent on geometry, and is computed by building probability maps for each PMT, $P_{Di}(x,y,z)$, using Geant4 simulations of optical photons generated at random throughout the vessel and then propagated through the liquid scintillator. 
   \item The efficiency of converting a photon into a photoelectron/PMT signal. This is defined on each PMT as the QE$_i$, or quantum efficiency, which may differ from PMT-to-PMT.
\end{enumerate}
The probability that a photon generated at a given position is detected successfully is given by QE$_i\times P_{Di}(x,y,z)$. Since this is a binary process, we utilise a binomial distribution to model the chance that $k$ photons are detected when $n$ photons have been generated,
\begin{equation}
   {\rm Bi}_i(k;n,x,y,z) =\begin{pmatrix}
 n\\ k
\end{pmatrix}\left({\rm QE}_i\times P_{Di}(x,y,z)\right)^k\left(1-{\rm QE}_i\times P_{Di}(x,y,z)\right)^{n-k}.
\end{equation}
The probability distribution for $k$ successful detections at PMT$_i$ given an energy deposition of $E$ at any position is given by
\begin{equation}
   {\rm Pr}_i(k;E,x,y,z) = \int_0^{\infty} {\rm Bi}_i(k;n,x,y,z) \times {\rm Pois}(n;{\rm LY}\times E) ~dn.
\end{equation}

Finally, we sum over each of the PMT probabilities defined above to find the total number of detected photons (which is otherwise the probability of detection at any PMT). This gives the following: 
\begin{equation}
\begin{split}
   {\rm Bi}_T(k;n,x,y,z) =\begin{pmatrix}
 n\\ k
\end{pmatrix}\left(\sum_{i}{\rm QE}_i\times P_{Di}(x,y,z)\right)^k&\left(1-\sum_{i}{\rm QE}_i\times P_{Di}(x,y,z)\right)^{n-k},\\
{\rm Pr}_T(k;E,x,y,z) = \int_0^{\infty} {\rm Bi}_T(k;n,x,y,z) \times &{\rm Pois}(n;{\rm LY}\times E) ~dn.
\end{split}
\end{equation}

The outcomes of these simulations are used to understand the detection efficiency of the liquid scintillator system, and set performance requirements.

\subsection{Performance Requirements}
\begin{figure}[!h]
   \centering
   \includegraphics[width=0.6\textwidth]{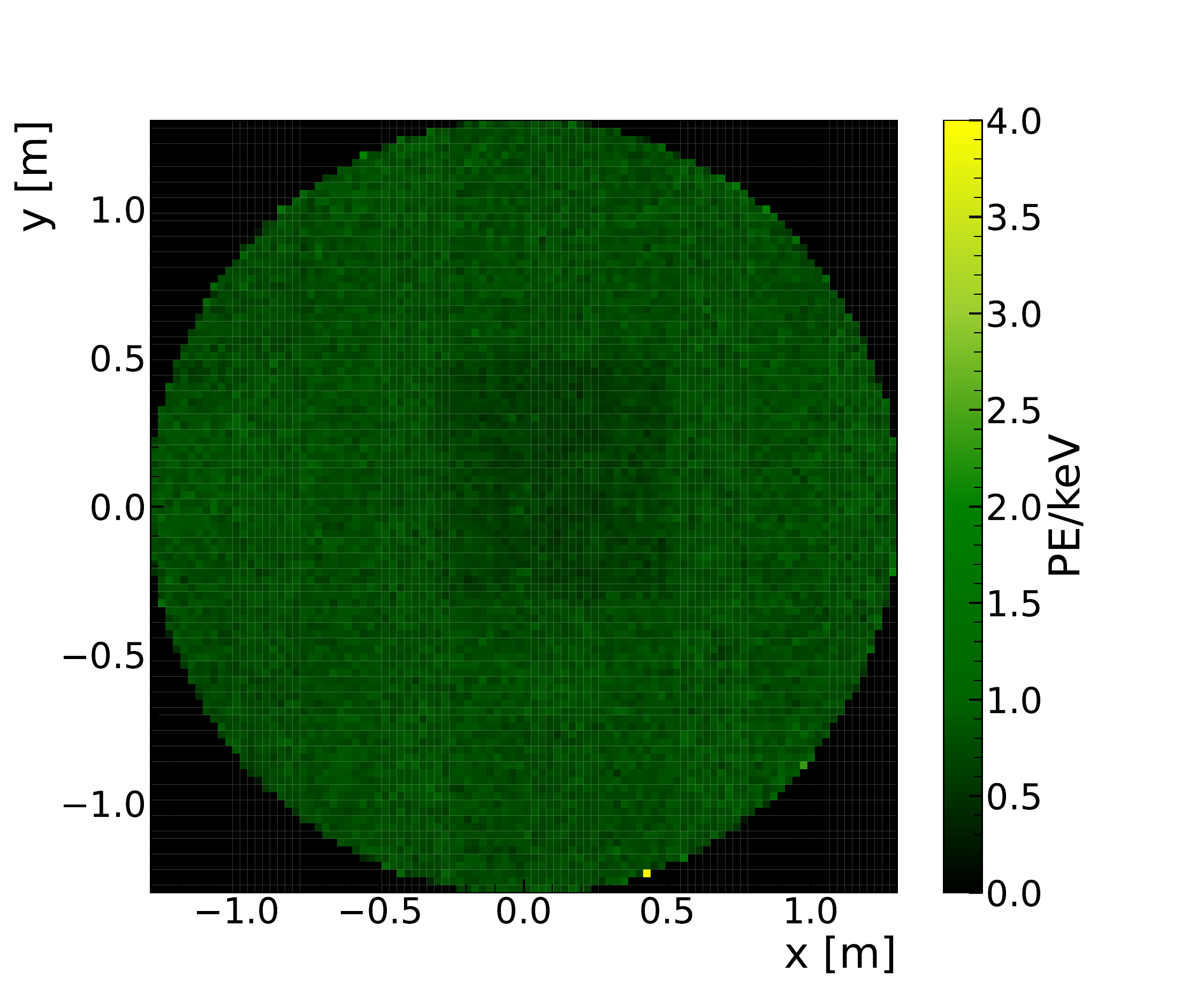}
   \caption{A two-dimensional histogram plotting the expected number of PE observed by any PMT, projected into the XY plane at Z=0 mm. There is lower probability of detection for photons produced between the crystal enclosures.}
   \label{fig:eff_pos_dep}
\end{figure}

The $P_{Di}(x,y,z)$ distribution is computed by randomly generating a large number of optical photons throughout the vessel, recording their initial position and whether or not they are incident on a PMT photocathode. Such a hit map is generated for every PMT. These can be combined to give the average probability of a single optical photon emitted anywhere in the vessel hitting any PMT, which is expected to be approximately 25\%. Assuming a veto PMT QE of 25\% \cite{veto-pmts}, and a light yield of 12 photons/keV, the expected average number of observed PEs is 0.75~PE/keV, or 0.042~PE/keV per PMT. This value depends on where the photons are generated because of the relatively sparse positioning of the PMTs compared to other large liquid scintillator detectors. An example of the expected position dependence of this is given in Fig. \ref{fig:eff_pos_dep}. It is worth noting the final positioning of PMTs is expected to be less sparse, with additional PMTs received from the decommissioned Daya Bay experiment bringing the total number of PMTs to 32. The final positioning is yet to be decided, and the PMTs are currently undergoing testing.

\begin{figure}[htb]
   \centering
   \includegraphics[width=0.49\textwidth]{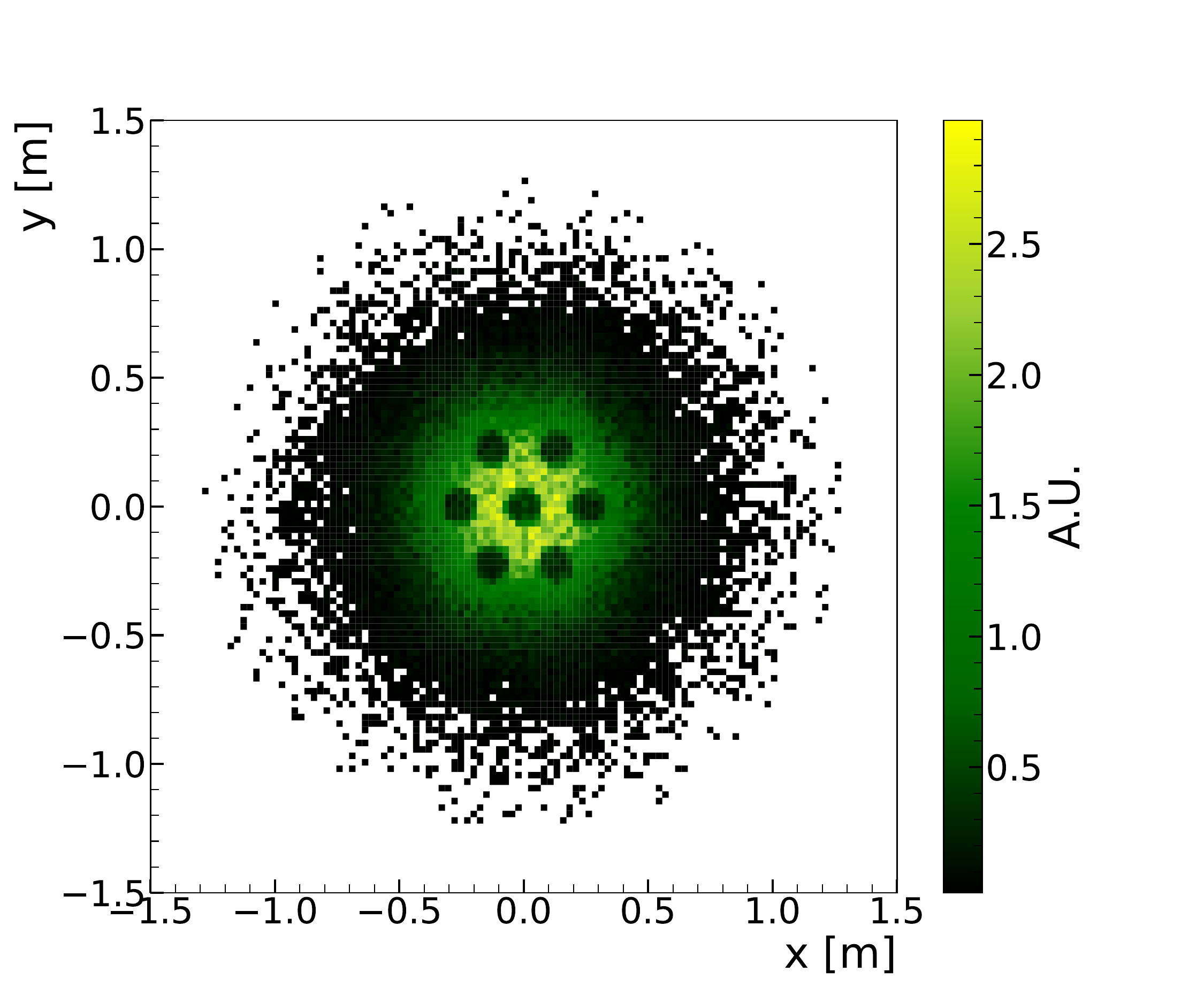}
   \includegraphics[width=0.49\textwidth]{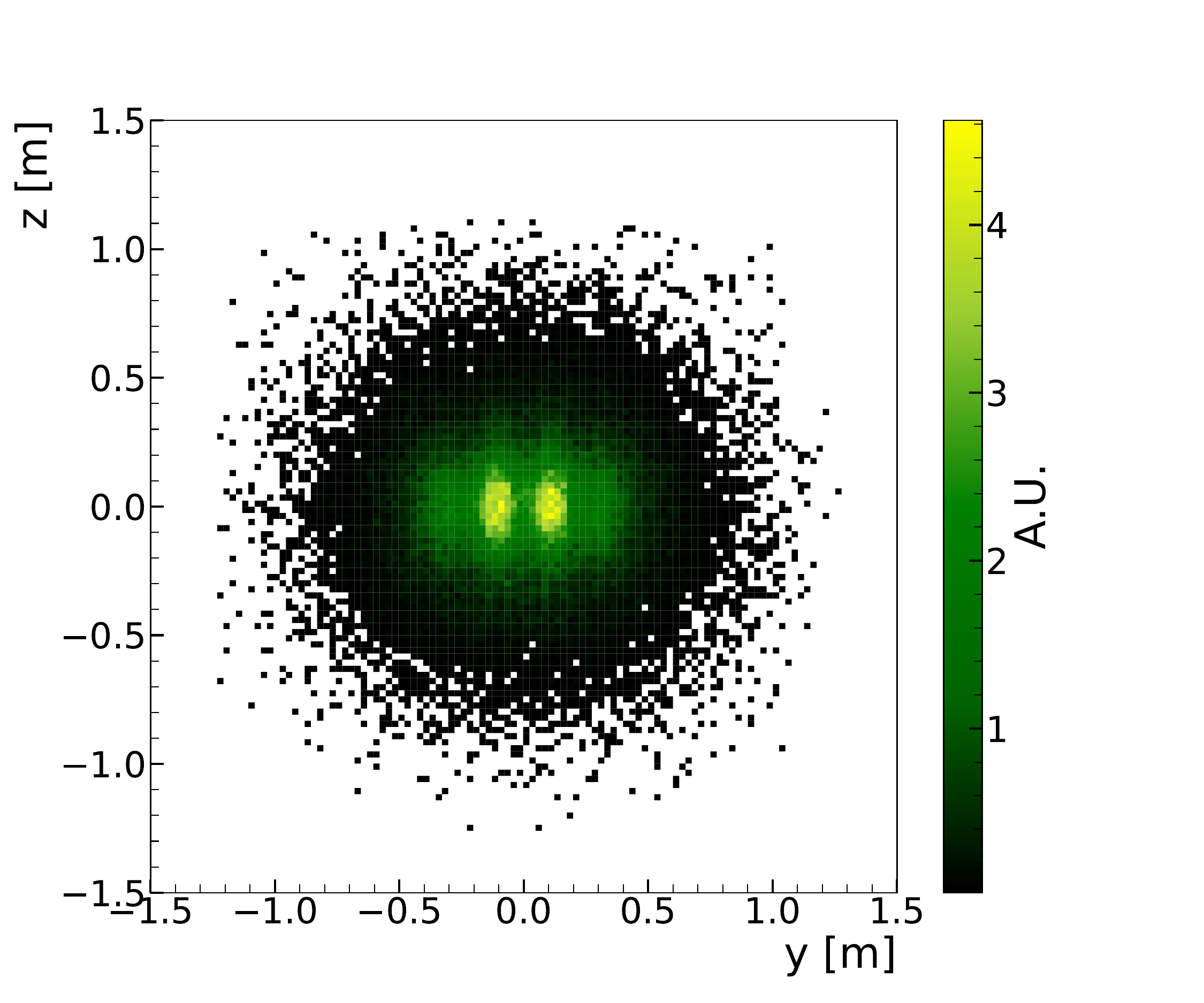}\\
   \includegraphics[width=0.49\textwidth]{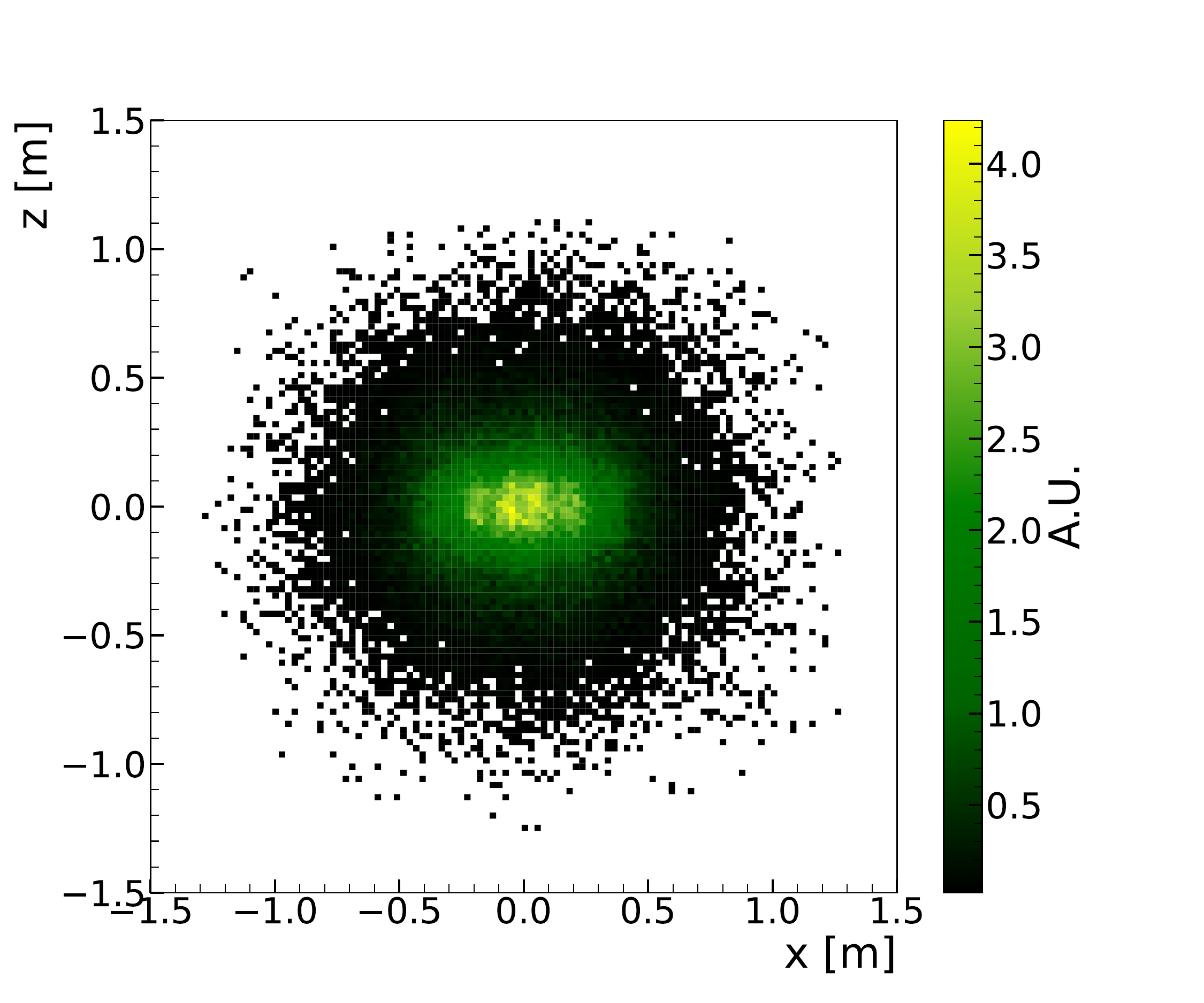}
   \includegraphics[width=0.5\textwidth]{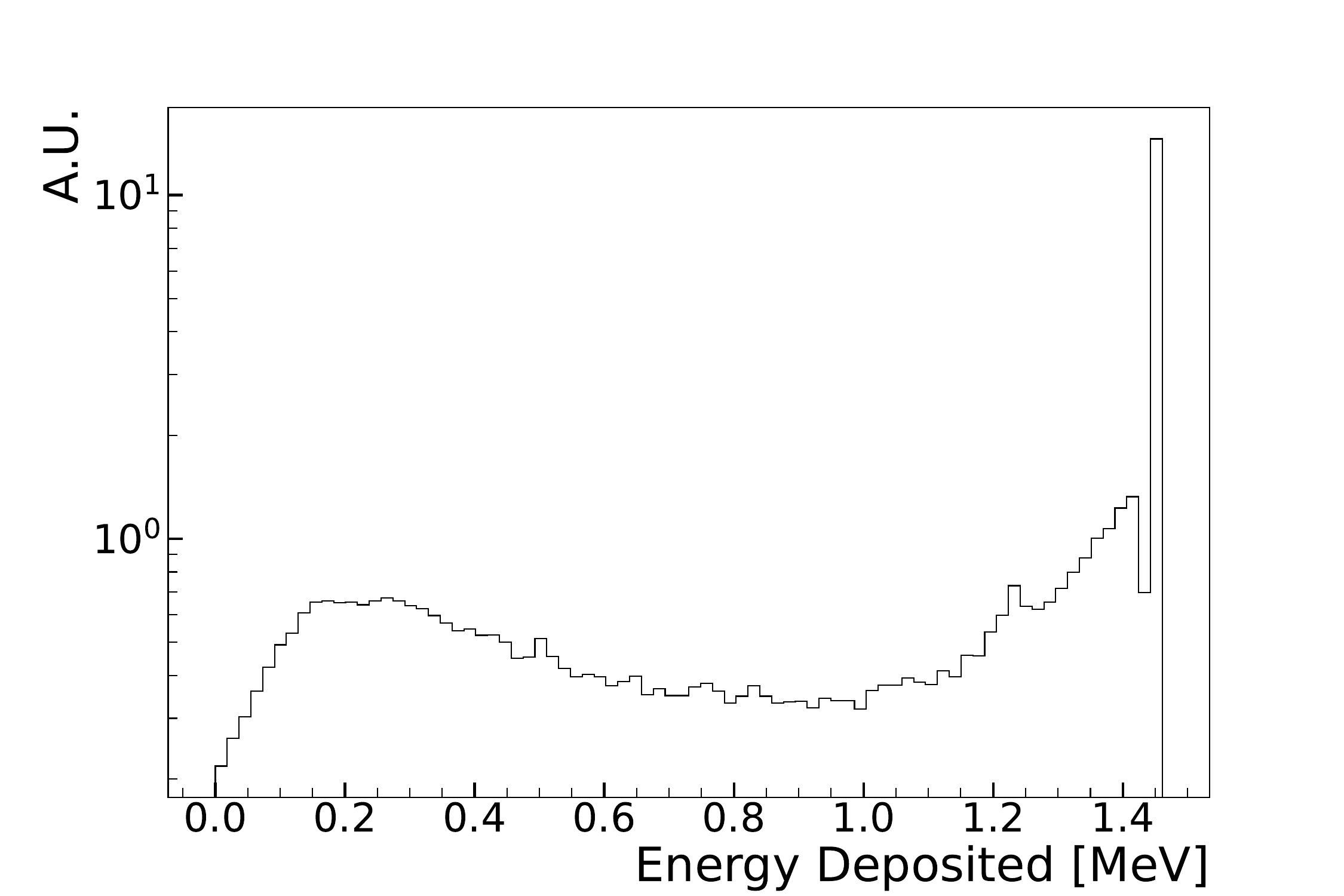}
   \caption{Distribution of crystal $^{40}$K background events in the liquid scintillator projected onto the XY (top left), YZ (top right) and XZ (bottom left) planes, and the energy distribution of deposition in the liquid scintillator (bottom right).}
   \label{fig:40kveto}
\end{figure}

Different threshold and coincidence requirements are used to study veto efficiency. The most important function of the veto is to detect background from $^{40}$K decays intrinsic to the crystals, and so we use this scenario to assess the performance of a number of different threshold requirements.
The distribution of hit positions in the liquid scintillator and the energy deposited for $^{40}$K decay events are shown in Fig. \ref{fig:40kveto}. The events are centred around the crystal array, where a large fraction of energy deposits are above 1 MeV, and are likely to be vetoed. There are still some events at or below 100 keV, where there may be inefficiencies.

\begin{figure}[htb]
   \centering
   \includegraphics[width=0.54\textwidth]{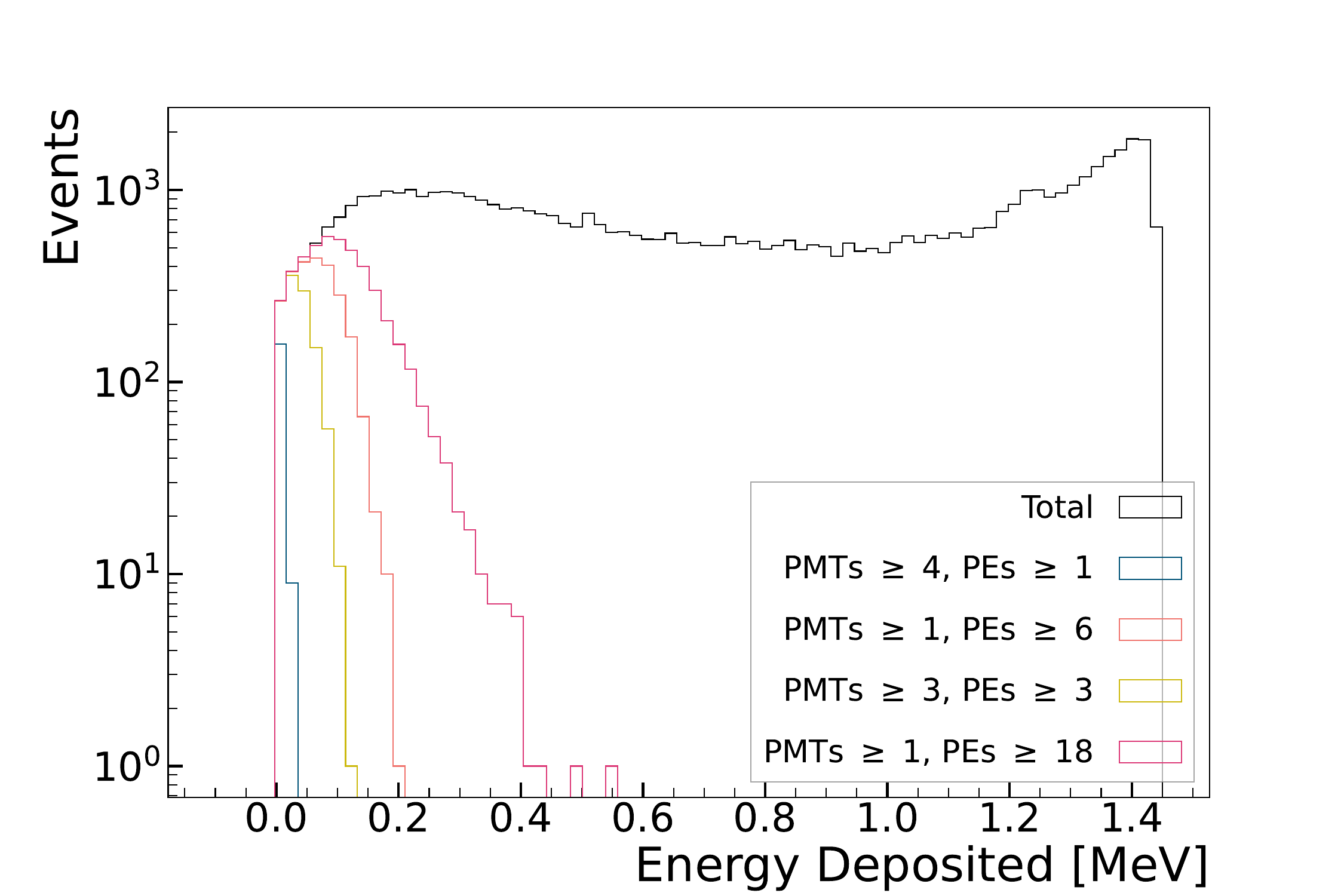}
   \includegraphics[width=0.45\textwidth]{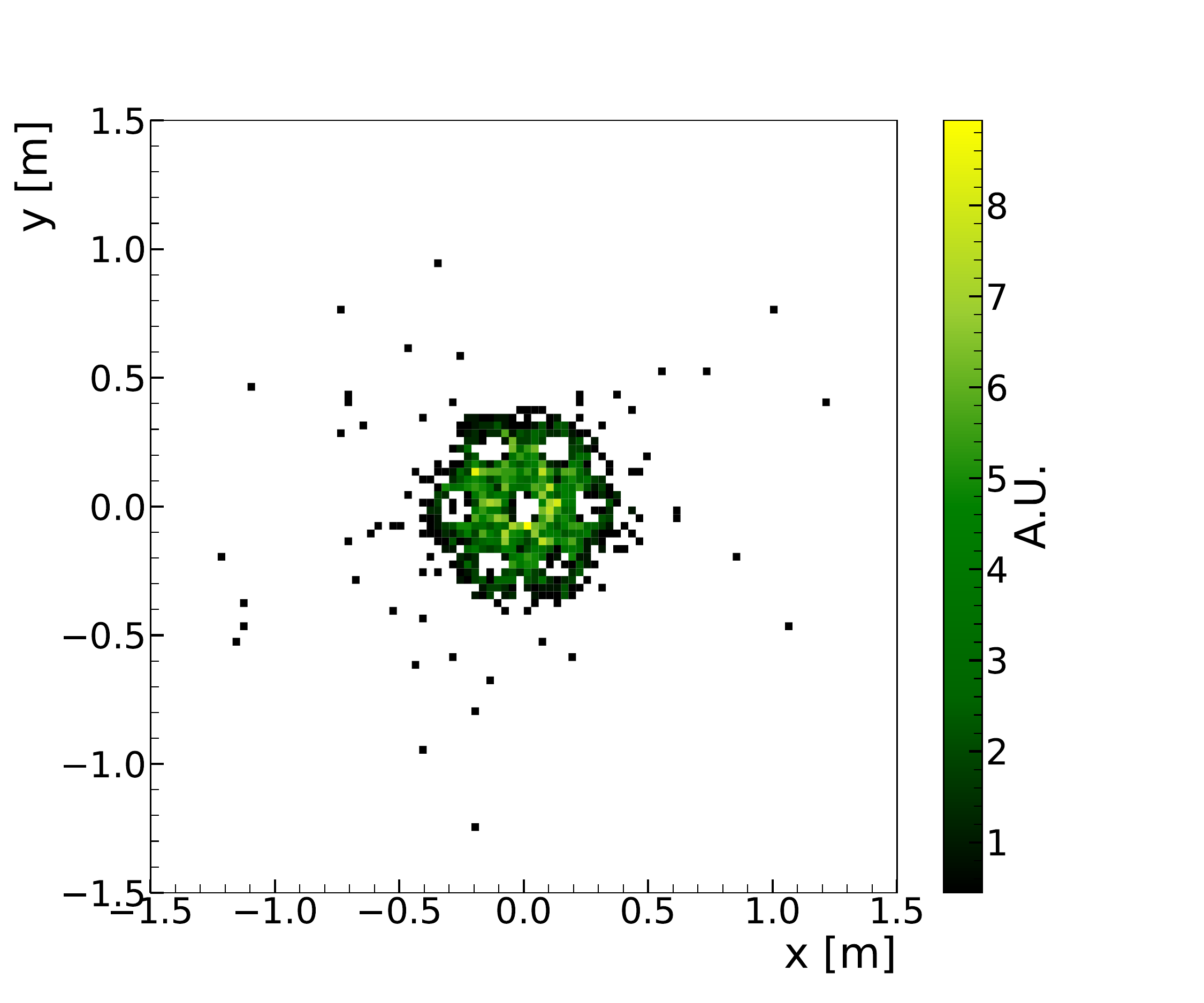}
   \caption{Left: energy deposited for events in liquid scintillator after the application of various veto threshold requirements. Right: the positions of said events after thresholds are applied. We can see that most events left after the application of PE/PMT based coincident requirements are below 200~keV, and are predominantly positioned in between the crystal enclosures.}
   \label{fig:veto-thresh}
\end{figure}

Veto efficiencies for the different threshold requirements are listed in Table~\ref{tab:thresh-eff}, and include efficiencies for the whole energy range, alongside those in some low energy bins. Most thresholds approach near perfect efficiencies above 100~keV. Proportions of events that survive each of a subset of these requirements are shown in Fig. \ref{fig:veto-thresh}, along with the position distribution of events that survive a 100~keV threshold.

\begin{table}[htb]
\centering
\caption{Detection efficiency in the liquid scintillator for $^{40}$K decays in the crystals that also deposit energy in the liquid scintillator. We present efficiencies for the total energy range, as well as for bins in the lower energy range. Note that not every event in the crystal will result in a coincident liquid scintillator deposition. The first column under threshold requirements is the minimum number of triggered PMTs, and the second is the minimum number of PEs required for a PMT to trigger.}
\begin{tabular}{ll|lllll}
\multicolumn{2}{c|}{Thres. requirement} & \multicolumn{5}{c}{Veto efficiencies} \\
\hline
Trig. PMTs  & PE Thres. & All energies & 0--20~keV & 20--50~keV & 50--80~keV & 80--100~keV\\ \hline
$\rm{PMTs} \geq 1$ & $\rm{PEs} \geq 1$        & 99.9\% & 92.5\% & 100.0\% & 100.0\% & 100.0\%      \\
$\rm{PMTs} \geq 4$ & $\rm{PEs} \geq 1$ & 99.9\% & 71.0\% & 100.0\% & 100.0\% & 100.0\%     \\
$\rm{PMTs} \geq 1$ & $\rm{PEs} \geq 6$           & 98.6\% & 0.0\% & 18.3\% & 72.6\% &  97.9\%    \\
$\rm{PMTs} \geq 3$ & $\rm{PEs} \geq 3$  & 99.3\% & 2.1\% & 67.3\% & 100.0\% & 100.0\%             \\
$\rm{PMTs} = 18$ & $\rm{PEs} \geq 1$    & 97.0\% & 0.0\% & 1.4\% & 25.2\% &  58.0\%            \\
\multicolumn{2}{c|}{$\rm{E} \geq 100$}    & 96.7\% & -- & -- & -- & --       \\
\multicolumn{2}{c|}{$\rm{E} \geq 50$}    & 98.7\% & -- & -- & -- & --      \\
\multicolumn{2}{c|}{$\rm{E} \geq 20$}    & 99.6\% & -- & -- & -- & --      \\ \hline
\end{tabular}
\label{tab:thresh-eff}
\end{table}
The events that remain after the application of various veto requirements are predominantly low energy depositions occurring in between the crystal enclosures (Fig.~\ref{fig:veto-thresh}). 

Evident from these results, is that in order to simultaneously maintain acceptable veto efficiencies and set low enough energy thresholds, sensitivity to single PEs across multiple PMTs is required. Being able to set any of these threshold requirements motivates a detailed understanding of the PMTs. Thus, each PMT must be calibrated, from both a charge response and noise perspective. The following two sections describe the pre-calibration of the PMTs to be used in the liquid scintillator veto system.

\section{The Hamamatsu R5912 PMT}

\begin{figure}[htb]
    \centering
    \includegraphics[width=0.85\textwidth]{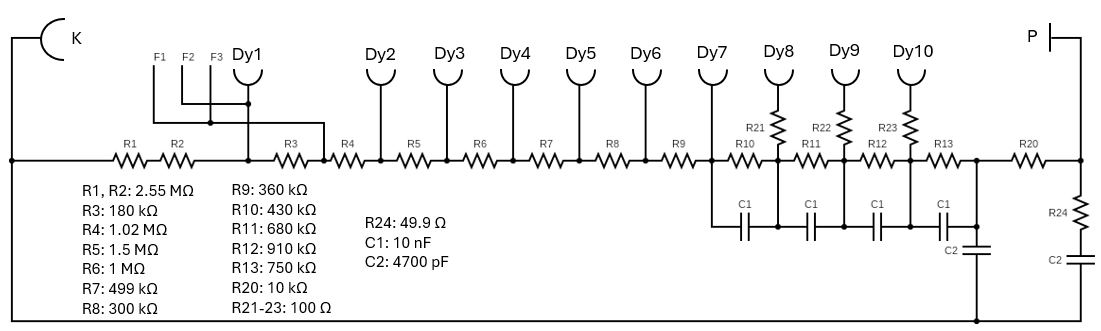}
    \includegraphics[width=0.485\textwidth]{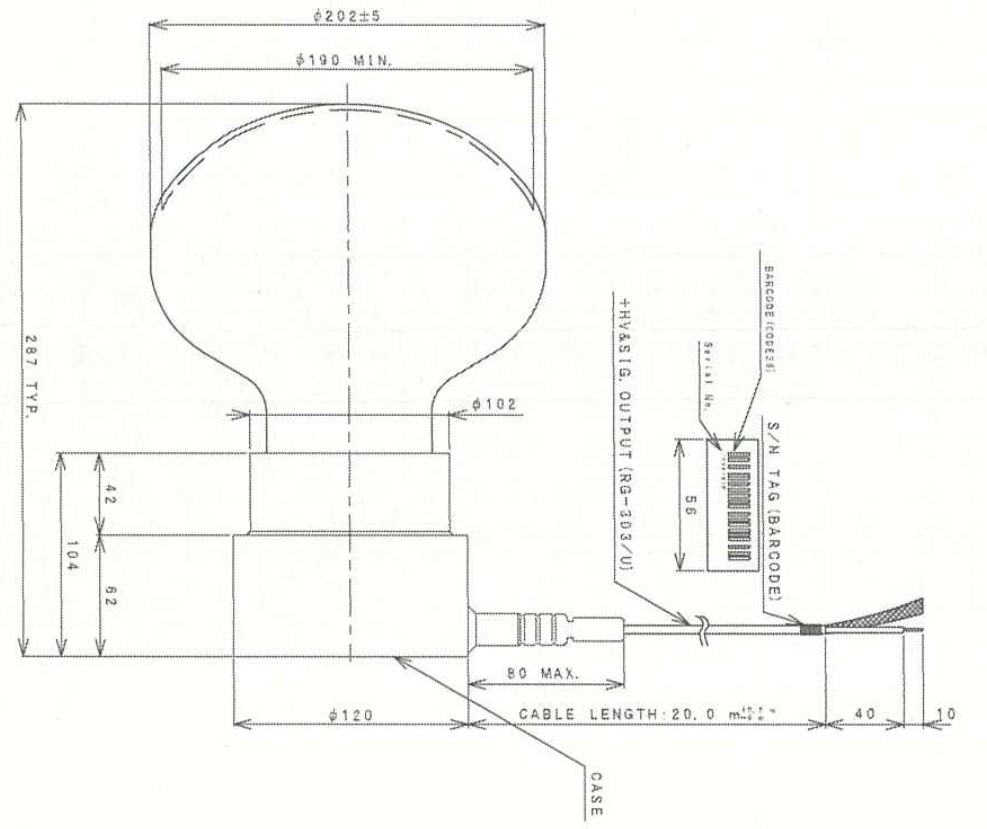}
    \caption{Top: A circuit diagram of the oil-proof base in the Hamamatsu R5912 PMTs. Bottom: A mechanical diagram of the R5912 PMT.}
    \label{fig:circuits}
\end{figure}

The PMTs are supplied from Hamamatsu with base electronics encased in an epoxy resin that is housed within an acrylic cylinder. This makes them sufficiently oil-proof for full immersion in the liquid scintillator. The circuit diagram for the supplied PMT base is shown in Fig.~\ref{fig:circuits}. The cabling comes fully attached to the potted base, and provides a single coaxial connection that requires decoupling of the HV and signal externally to the liquid scintillator vessel. Reference~\cite{veto-pmts} outlines some of the mechanical and response specifications provided by the manufacturer of the R5912 PMT, with a diagram of the PMT shown in Fig.~\ref{fig:circuits}. Bulk characterisation of these PMTs will need to include the gain and mean single photoelectron (SPE) charge determination, so that detector threshold conditions can be judiciously selected to achieve optimal photon detection efficiency and veto efficiency. Relative quantum efficiency measurements are performed to understand the detection efficiency, alongside timing and response linearity measurements, in order to inform reconstruction of background processes in the veto sub-detector. Temperature-dependent dark rate measurements are conducted to quantify noise levels, establish appropriate threshold conditions, and understand potential time-dependent background variations caused by temperature fluctuations.

\section{PMT Pre-calibration}

\subsection{Measurement Setup}
\label{sec:setup}

All optical measurements (excluding the temperature dependent dark rate study) are performed on a Thorlabs optical bench, using a Hamamatsu PLP-10 pico-second pulsed laser driver with a 405~nm laser head~\cite{PLP_10Laser}. The setup is housed in a large aluminium dark box mounted on the optical bench. The measurements performed in the dark box are at a laboratory temperature of 22~$^\circ$C and are monitored with RTD sensors to ensure that they do not vary by more than 1~$^\circ$C. 
The laser is routed into the dark box via 200~$\mu$m core multimode optical fibre patch cable, collimated, and then attenuated through a pinhole and neutral density (ND) filter before being detected by the PMT, or split via a reflective beam-splitter and then subsequently detected.


The PMTs are biased using a CAEN A7435~\cite{CAEN_7435} high voltage (HV) power supply mounted in a CAEN SY5527 HV mainframe~\cite{CAEN_5527}, with voltage control provided by the SABRE DAQ software. Waveforms are recorded with a CAEN V1730D~\cite{CAEN_1730} digitiser, with a sampling rate of $500\, \mathrm{MS/s}$, $0.12$ mV resolution, and a dynamic range of 2 V$_{\mathrm{pp}}$. These are the nominal digitisers for the SABRE experiment.
Data acquisition is managed by SABRE South experiment DAQ software~\cite{TDRSUmm}. Files are output in binary format and then processed and analysed in Pyrate~\cite{Scutti:2023pyz}, a python based analysis framework developed for the SABRE experiment. For the experiments conducted in this paper, the CAEN Raw firmware~\cite{CAEN_RAW} is used for acquisition of bulk SPE/gain data, and the PSD (pulse shape discrimination) firmware~\cite{CAEN_PSD} is used for the temperature dependent dark rate measurements. The raw firmware is chosen for the SPE measurements due to its more general trigger implementation, as it will acquire all channels when one channel is triggered, whereas the PSD can be configured to collect data only for an individual channel when it is triggered. Given the dark rate measurements only acquired one channel at a time, the PSD firmware was chosen over the raw due to its treatment of the baseline. When setting thresholds and deciding when to trigger, the PSD firmware will compare any threshold to a baseline calculated individually for that event. This eliminates the possibility of any baseline drift impacting the trigger rate, and therefore the dark rate measurement.

\subsection{Single Photo-Electron Response and Gain}
\label{sec:SPE}

The SPE response measured in this study is the charge distribution of the PMT pulse, specifically the integration of the region of the waveform containing the PMT pulse. Signal pulses are identified using a leading edge threshold algorithm applied during offline analysis. The pulse window is defined relative to a 1.8~mV threshold crossing point, extending from $10$ ns before the crossing point and $50$ ns afterward. The charge spectrum resulting from the integration of this pulse window for each event is used to determine the gain. This is achieved by fitting said spectra and determining the single photo-electron charge for a range of bias voltages.

\begin{figure}[htb]
    \centering
    \includegraphics[width=0.9\textwidth]{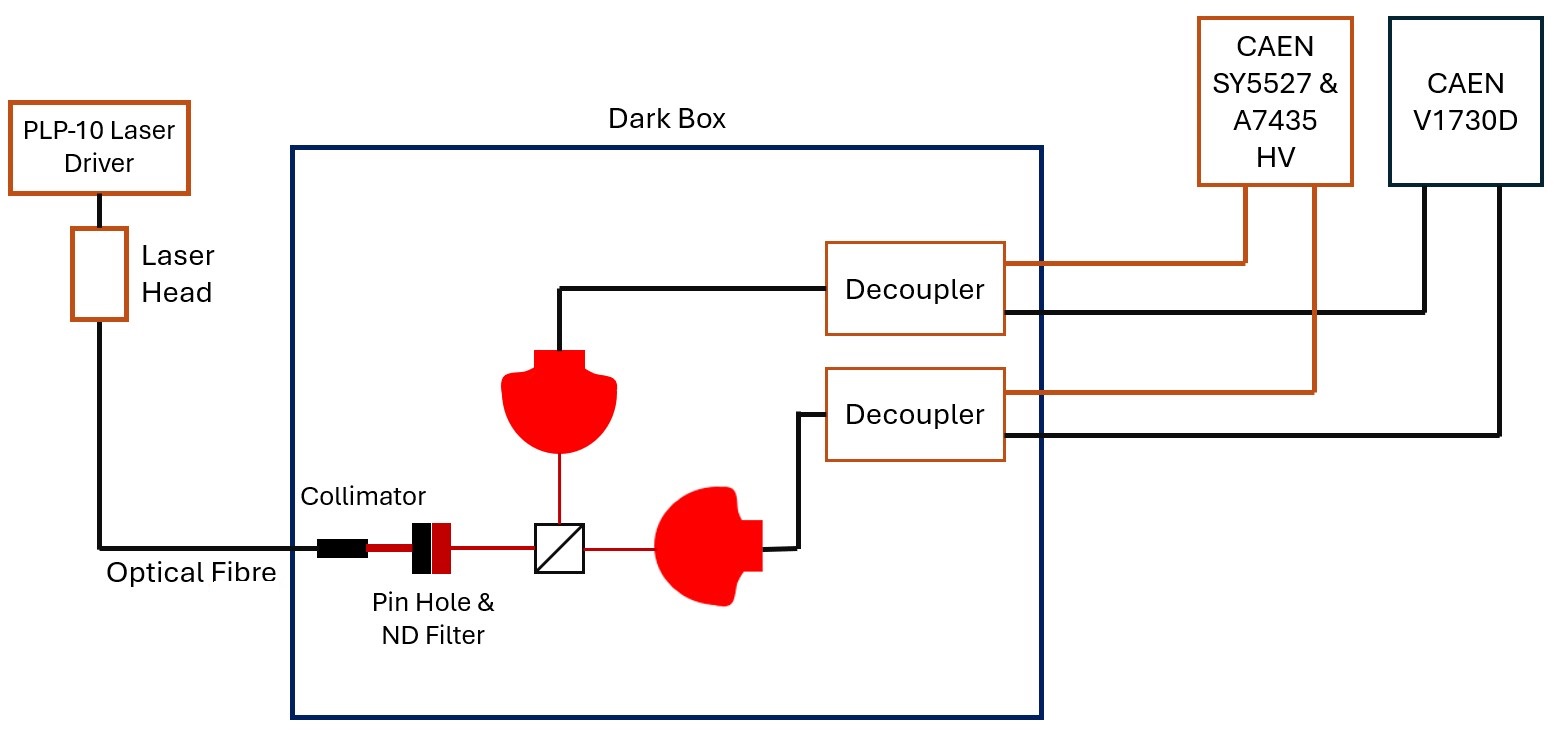}
    \caption{A diagram of the optical setup and electronics used for SPE measurements. These electronics are also used for the other measurements.}
    \label{fig:SPESetup}
\end{figure}

The SPE response of each PMT is measured via the optical setup described in Sec.~\ref{sec:setup}, which is capable of producing low enough pulse occupancies such that, statistically, most detected pulses contain a single photon. This is achieved by setting the pulsed laser to pulse at a frequency of 10~kHz, and that the attenuation of the ND filter chosen is $10^4$. Two PMTs are tested at once, so a beamsplitter is used after the ND filter. A diagram of this setup is shown in Fig.~\ref{fig:SPESetup}.

Both the PMT and the laser signals are acquired with the V1730 digitiser. The DAQ is configured to trigger on events in either PMT, thus only acquiring laser data when the PMT triggers, with a threshold of 1.83~mV (15~ADC). We chose these trigger settings to reduce electronic noise in the dataset, and ensure laser signals from each PMT were acquired. Single photo-electron events are selected by requiring a coincidence with the laser trigger signal. The timings of each pulse are determined using constant fraction discrimination (CFD). A time difference is determined between the signals from the PMTs and the laser and a 20 ns window is selected around the peak of the distribution that corresponds to the time difference between the pulse emission and subsequent detection in the PMT.

\begin{figure}[htb]
    \centering
    \includegraphics[width=1\textwidth]{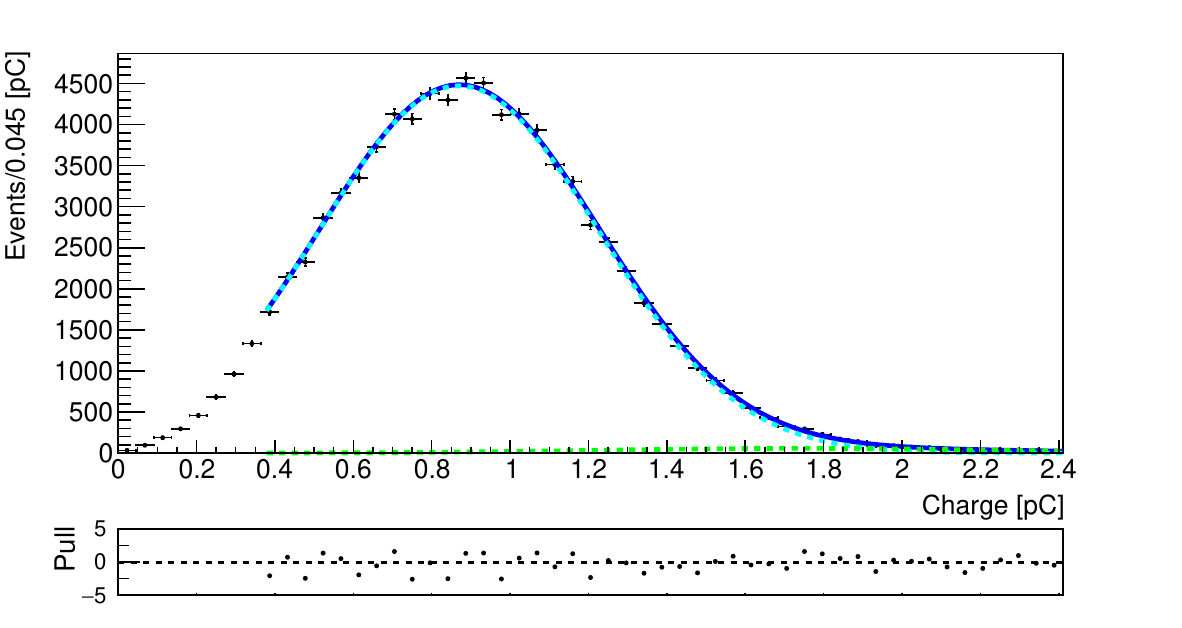}
    \caption{An example of an successful fit used to extract an SPE charge value, in linear scale. This is an example at 1500~V, with the fit results $\mu_{\rm1PE} = 0.870 \pm 0.002$~pC and $\sigma_{\rm1PE} = 0.36$~pC. This spectrum was fit using RooFit~\cite{RooFit}.}
    \label{fig:SPEFit}
\end{figure}

\begin{figure}[htb]
    \centering
    \includegraphics[width=0.495\textwidth]{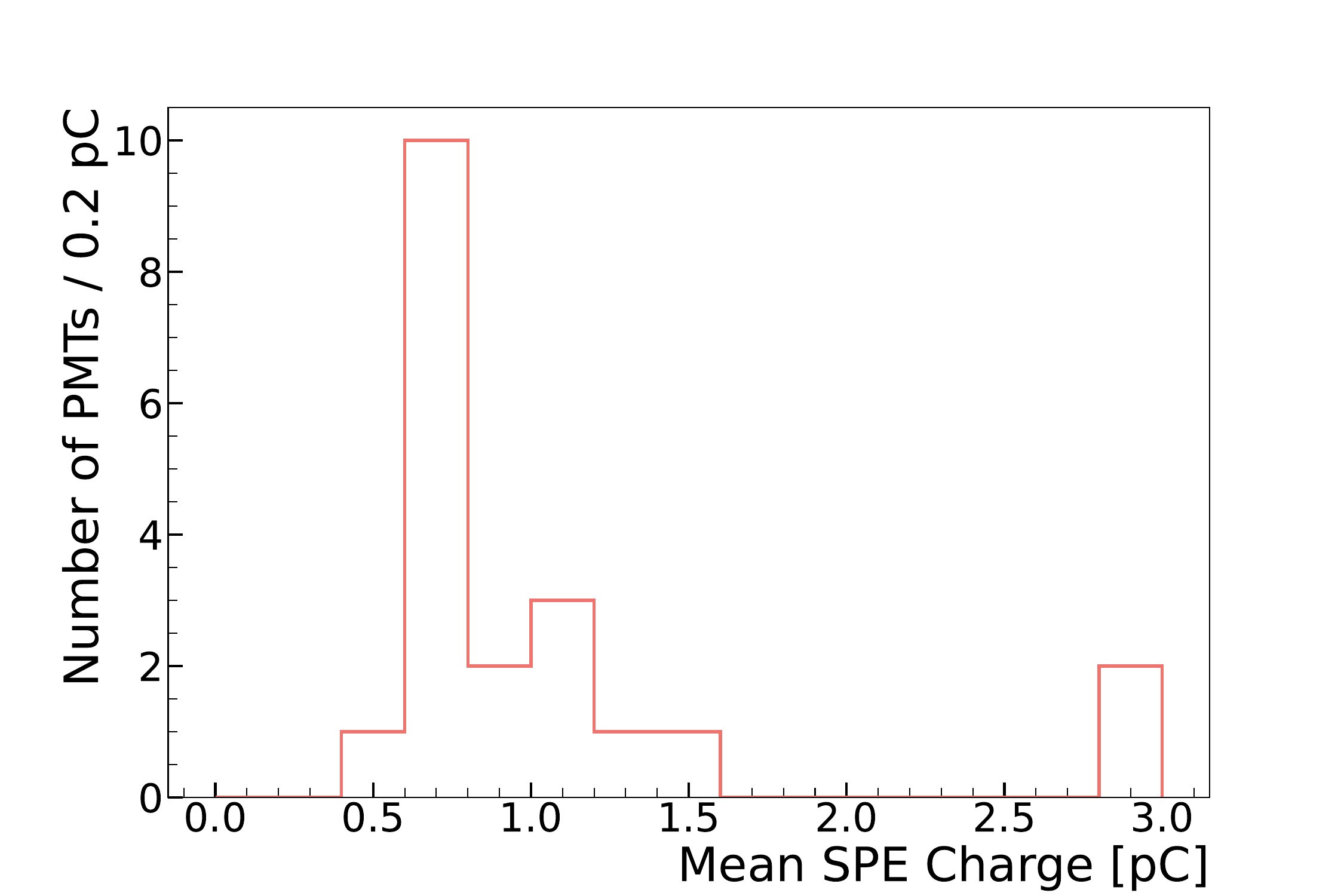}
    \includegraphics[width=0.495\textwidth]{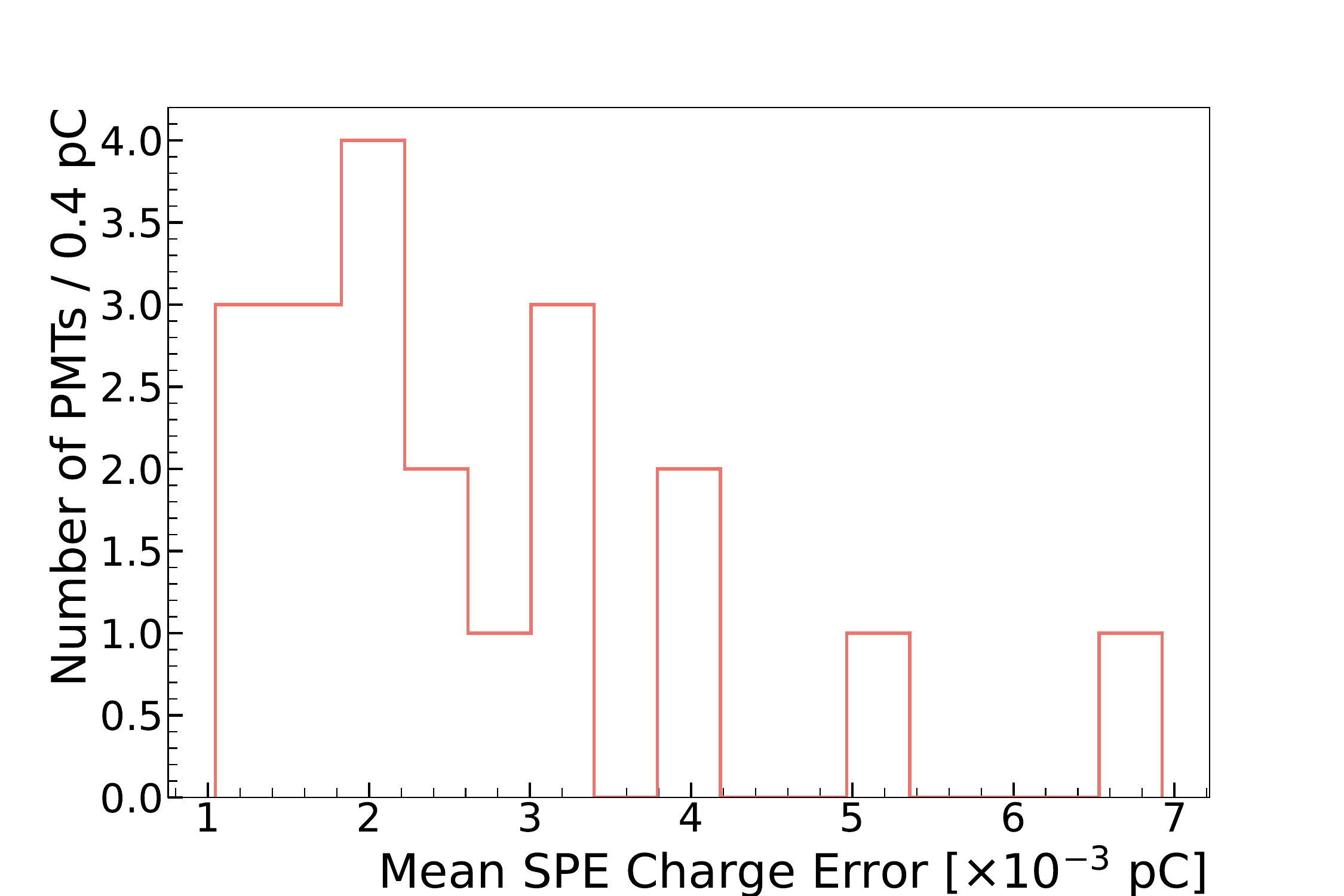}
    \includegraphics[width=0.495\textwidth]{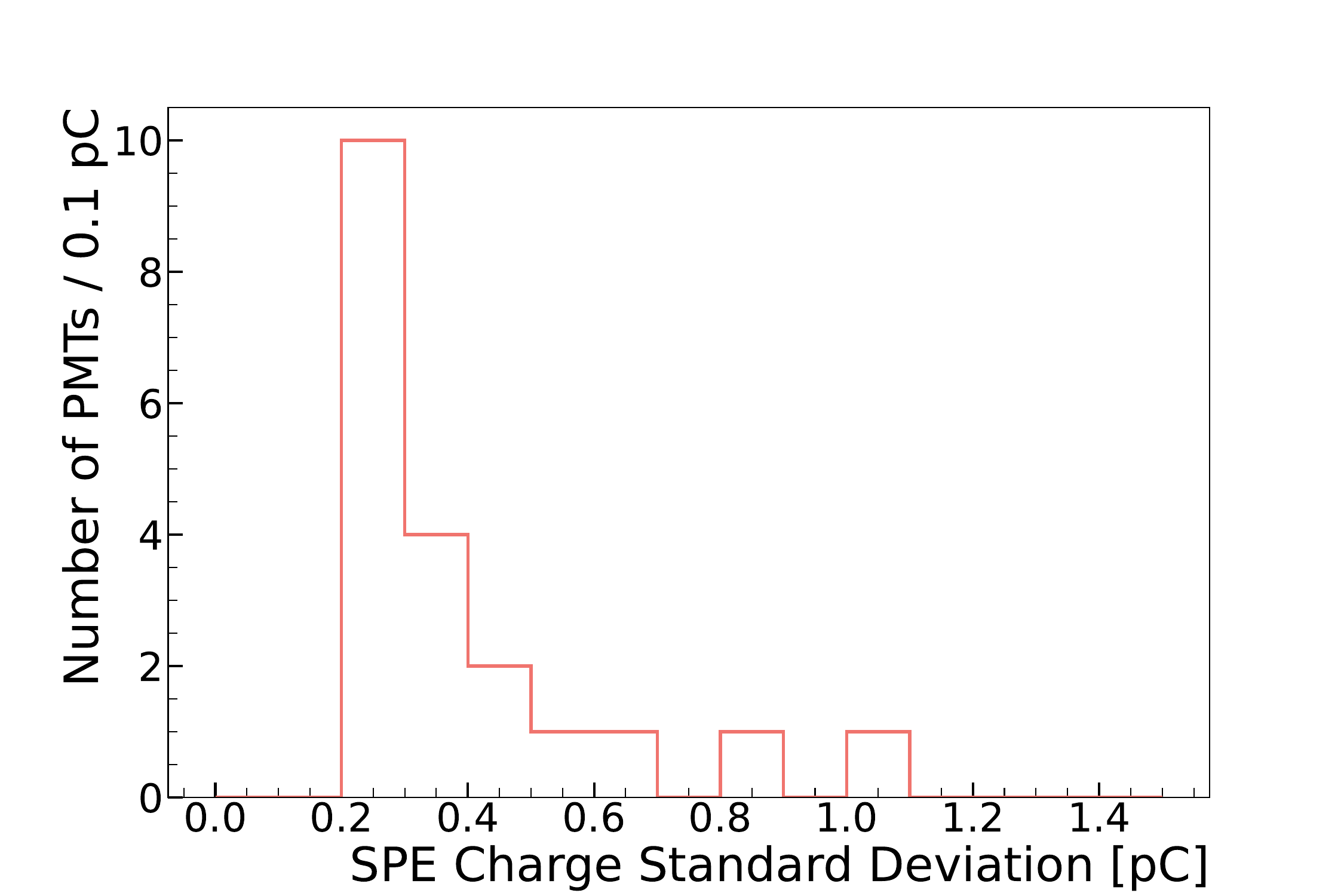}
    \caption{Values of $\mu_{\rm1PE}$ for all 20 R5912 PMTs that were tested, at a fixed bias voltage of 1500~V. There are two outliers with higher than average gains, with SPE charges just under 3~pC. Also included is the error on $\mu_{\rm1PE}$, and values for $\sigma_{\rm1PE}$}
    \label{fig:BulkSPERes}
\end{figure}

The DAQ threshold introduces a trigger bias at low charges. Figure~\ref{fig:FullModelFits} shows charge spectra without this DAQ threshold. The trigger bias means that low charge components cannot be modelled, such as the pedestal and any underamplified contributions (said contributions are explored in Sec.~\ref{sec:fullModel}).  As a result, a simplified model is used for the purpose of bulk estimation of the SPE charge, defined as the sum of two weighted Gaussians:
\begin{equation}
   F(N_{\rm1PE}, \mu_{\rm1PE}, \sigma_{\rm1PE}) = N_{\rm1PE} G(\mu_{\rm1PE}, \sigma_{\rm1PE}) + (1-N_{\rm1PE})G(2 \mu_{\rm1PE}, \sqrt{2} \sigma_{\rm1PE})
\end{equation}
where $G(\mu, \sigma)$ represents a standard Gaussian, and $N_{\rm 1PE}$ the number of SPEs. The first Gaussian models the 1 PE charge distribution, and the second the 2 PE charge distribution, hence, $\mu_{\rm 2PE} = 2 \times \mu_{\rm 1PE}$ and $\sigma_{\rm 2PE} = \sqrt{2} \times \sigma_{\rm 1PE}$. The fit range is defined to begin above the region that is affected by the trigger bias. This starting point is custom for each fit and is determined by scanning a range of starting values and picking the value that best optimises the fit $\chi^2/NDF$. To ensure the quality of these fits, $\chi^2/NDF$ and $\mu_{1PE}$ are required to be stable within the range from which the fit start value is chosen. An example of such a fit is shown in Fig.~\ref{fig:SPEFit}. The results from all PMTs are summarised in Fig.~\ref{fig:BulkSPERes}, which shows the values of $\mu_{\rm1PE}$ at the nominal bias voltage of 1500~V, along with values for the error on $\mu_{\rm1PE}$, and values for $\sigma_{\rm1PE}$. 

\begin{figure}[htb]
    \centering
    \includegraphics[width=0.495\textwidth]{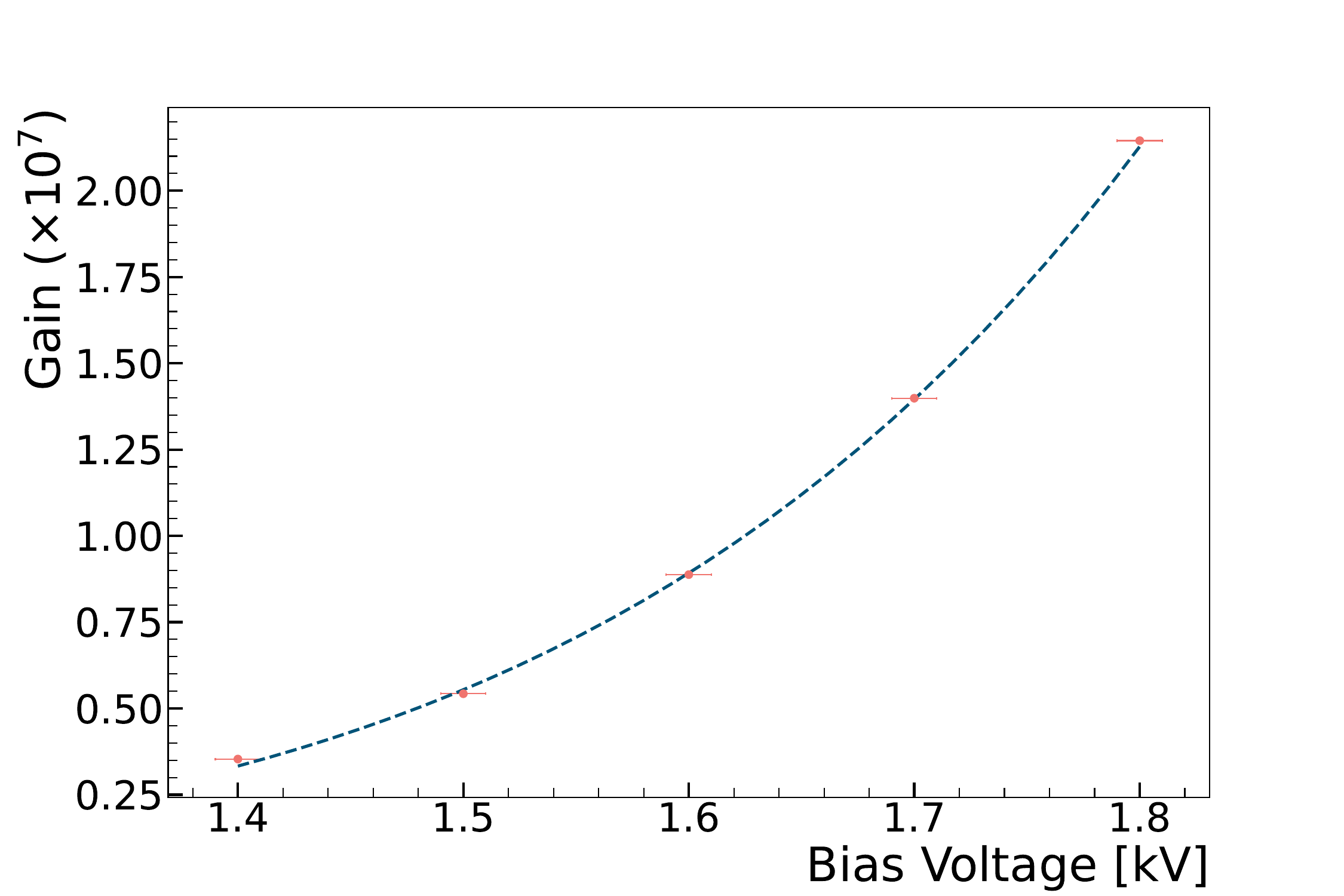}
    \includegraphics[width=0.495\textwidth]{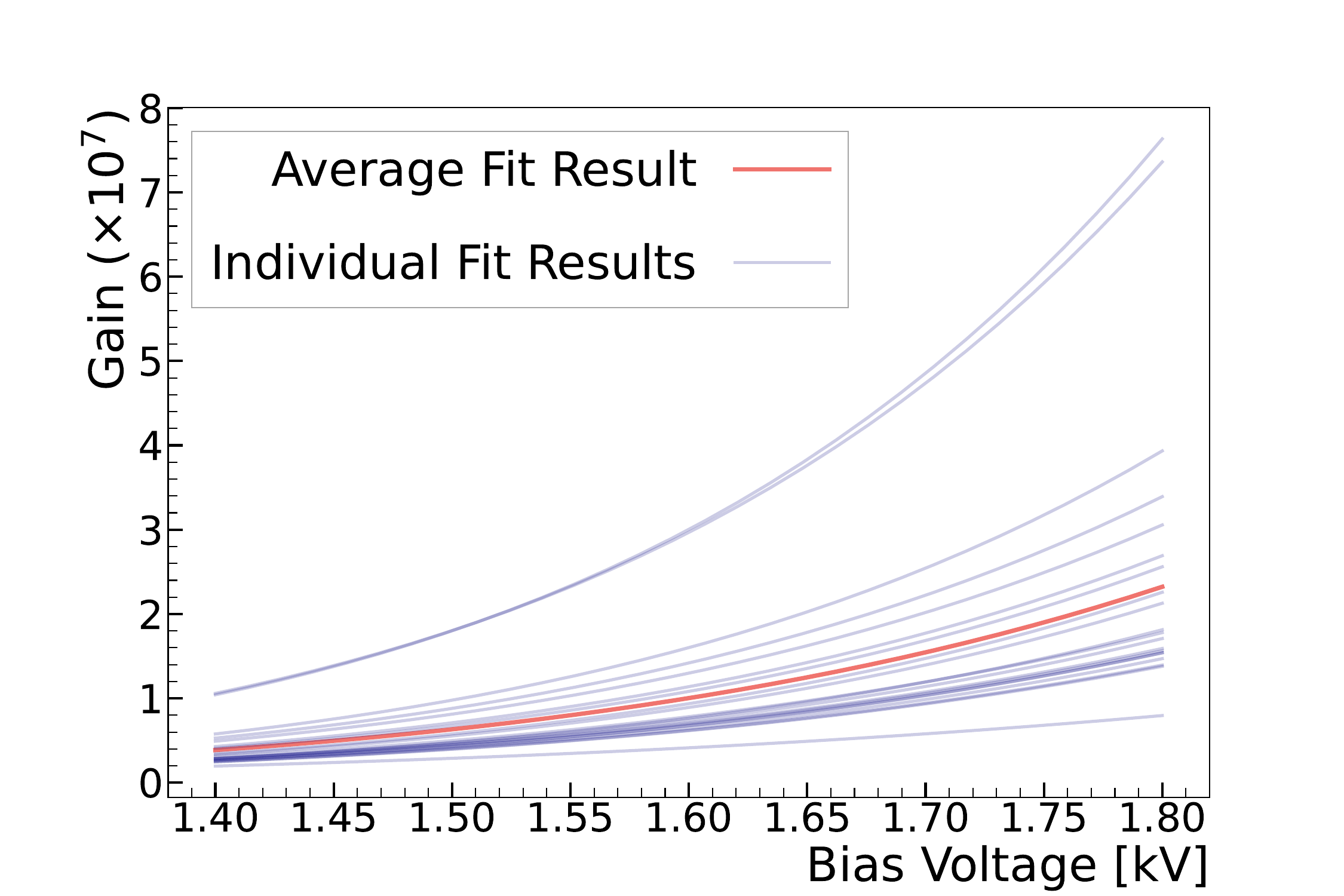}
    \caption{Right: The curve for $\mu_{\rm gain}$ for one of the SABRE South R5912 PMTs, fit over a range of 1400~V to 1800~V. Left: $\mu_{\rm gain}$ curves for each of the calibrated PMTs, as well as the average curve.}
    \label{fig:GainFit}
\end{figure}

For each PMT and at each bias voltage, we can take the ratio of the SPE charge with the electron charge to determine the gain. The gain as a function of voltage can be modelled with the expression (given by Hamamatsu~\cite{HamamatsuHandbook}): 
\begin{equation}
   \mu_{\rm gain} = AV^{kN}
\end{equation}
where $A$ and $k$ are floated parameters specific to a given PMT, and $N$ is the number of dynodes in the PMT --- in the case of the Hamamatsu R5912, $N=10$. The gain for each PMT is successfully modelled according to this expression, between the bias voltages of 1400~V and 1800~V in 100~V steps, using a fit defined via the lmfit package~\cite{lmfit}. An example of a fitted gain curve is shown in Fig.~\ref{fig:GainFit}, alongside a summary plot depicting the gain curves of each of the 20 calibrated PMTs and their average.

\subsubsection{Model of Response with Undersized Components}
\label{sec:fullModel}

\begin{figure}[htb]
    \centering
    \includegraphics[width=0.495\textwidth]{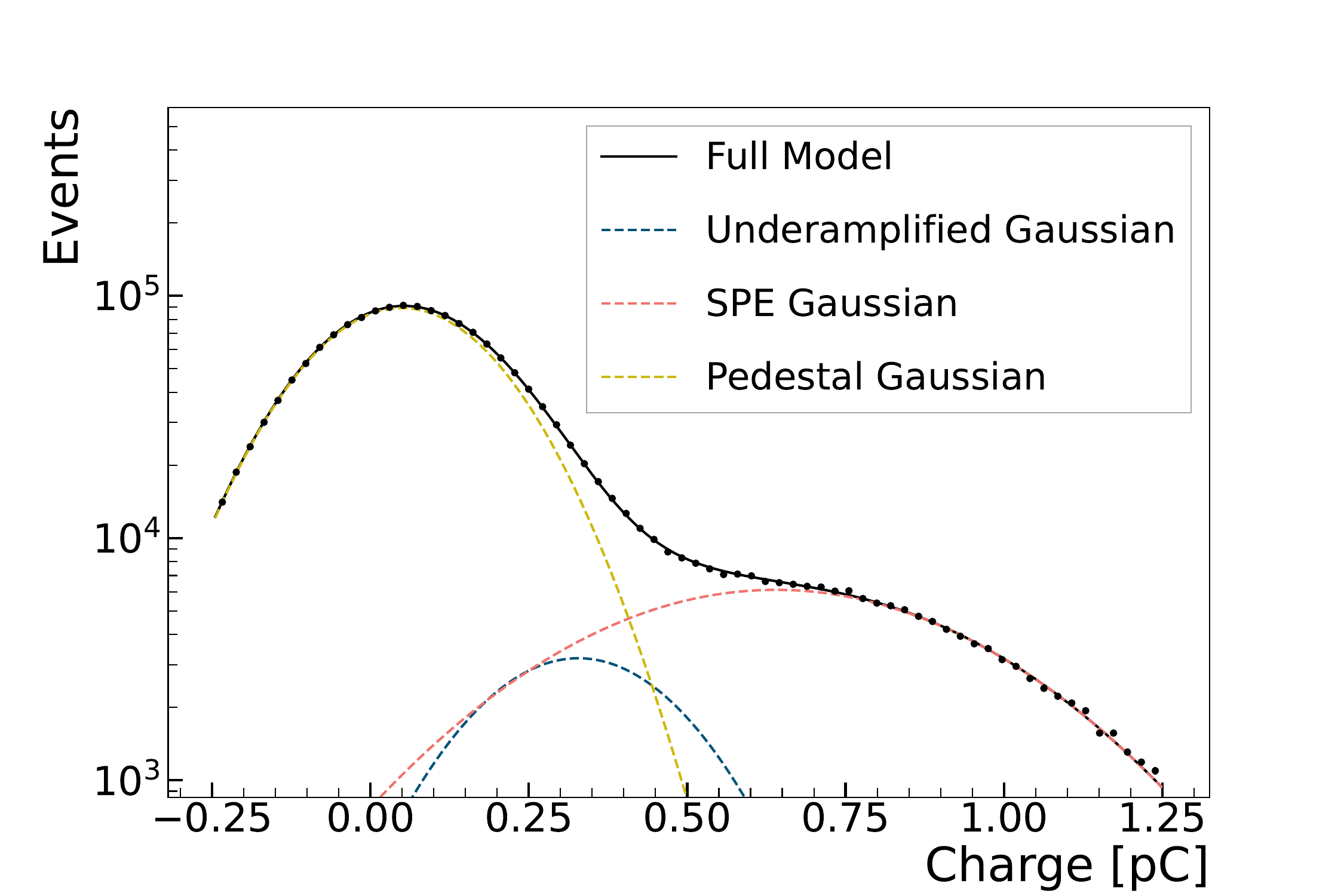}
    \includegraphics[width=0.495\textwidth]{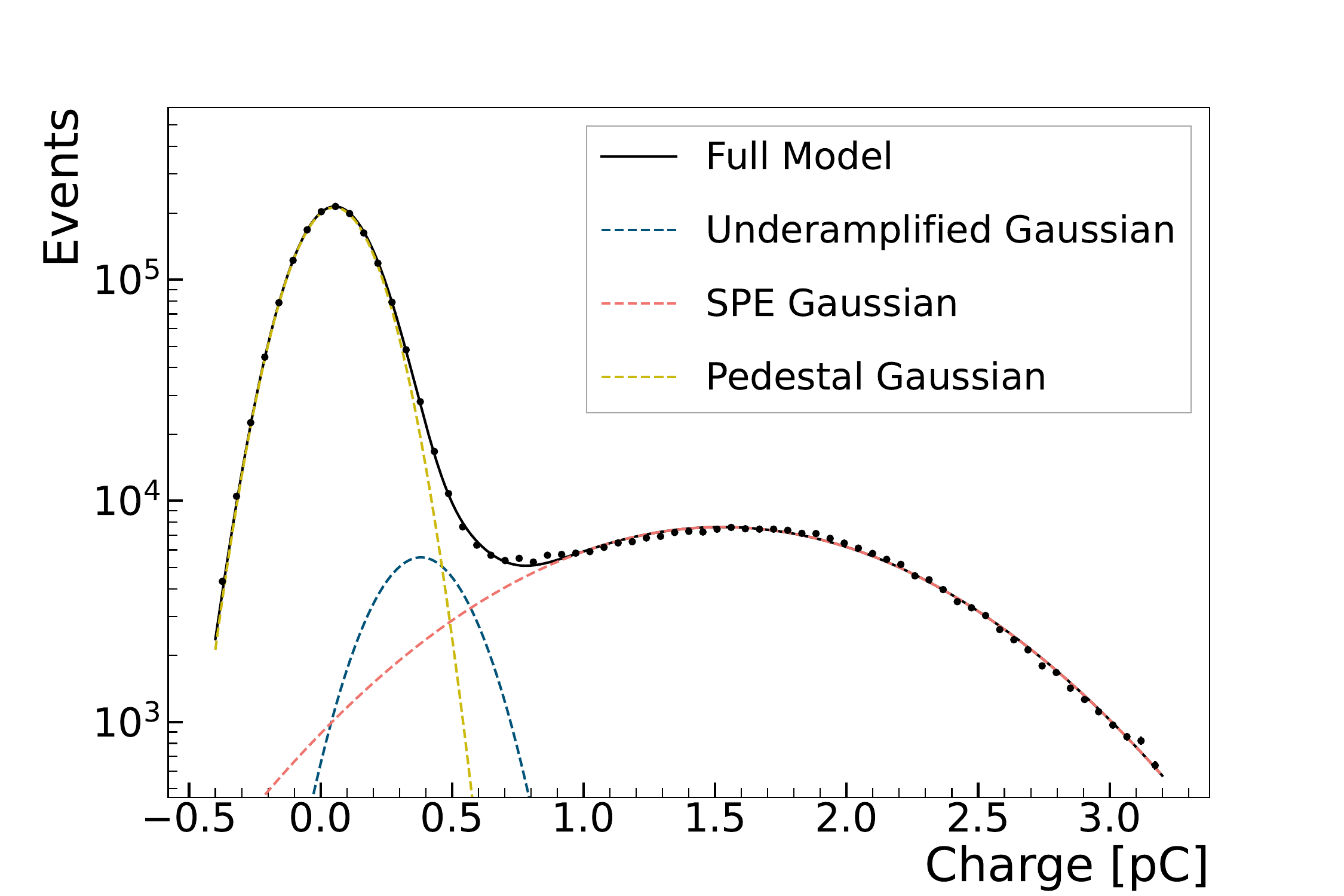}
    \caption{SPE models including a pedestal and an underamplified component, with 1500~V on the left and 1700~V on the right. Whilst higher PE components are not included, it is expected they should follow the parametrisation laid out in Sec.~\ref{sec:SPE}.}
    \label{fig:FullModelFits}
\end{figure}

A full measurement of the response of the PMTs at lower charges is necessary to provide a complete parameterisation of the SPE response for simulation and digitisation studies reproducing PMT pulses, for example, in particle ID studies that leverage pulse shape variables. To build such a model, we used a PMT with a gain and charge response that is representative of the overall PMT set. The single photoelectron (SPE) measurement was triggered by the laser, not the PMT itself. This approach avoids the need for a hardware threshold on the PMT and allows the pedestal to be included in the fit. This method could not be applied to the bulk estimation, as the data was collected via a hardware threshold on the PMT waveform. The laser triggered charge is calculated in a waveform window defined relative to an existing laser pulse, based on the expected location of the pulse from a PMT, with a width of 120~ns. The resulting charge spectrum was fit according to:
\begin{multline}
    F(x, N_1, N_2, \mu_{\rm 1PE}, \sigma_{\rm 1PE}, \mu_{\rm ped}, \sigma_{\rm ped}, \delta) = N_1 G(x, \mu_{\rm ped}, \sigma_{\rm ped})
    + N_2 G(x, \mu_{\rm 1PE}, \sigma_{\rm 1PE}) +  \\ (1 - N_1 - N_2)G(x, \delta \times \mu_{\rm 1PE}, \delta \times \sigma_{\rm 1PE}),
\end{multline}
where the parameters $N_1$ and $N_2$ are constants defined in the range [0,1] so to maintain the normalisation of the full PDF, as is the parameter $\delta$ so that the underamplified Gaussian can be properly constrained, $G(\mu, \sigma)$ is a Gaussian, and the remaining 4 parameters define the means and variances of the SPE Gaussian and pedestal. The model includes a Gaussian for the pedestal, the 1PE response, and for underamplified 1PE pulses, respectively. An extended fit performed with the zfit pythonic fitting package~\cite{zfit} is used to fit the PMT SPE responses at three bias voltages: 1500~V, 1600~V, and 1700~V. Figure~\ref{fig:FullModelFits} shows the 1500~V and 1700~V fit in increasing order of voltage. The above model fits each dataset well.

\subsection{Dark Rate}

Temperature-dependent measurements are performed at temperatures recorded to be within $1^{\circ}$C of 5 preset values: 293.15~K ($20^{\circ}$C), 295.65~K ($22.5^{\circ}$C), 298.15~K ($25^{\circ}$C), 300.65~K ($27.5^{\circ}$C), 303.15~K ($30^{\circ}$C), at bias voltages of corresponding to 1400~V, 1500~V, 1600~V and 1700~V. The operating temperature of SUPL will be between 295.65~K and 298.15~K, so testing in the above range should cover even the most drastic temperature variations.
The thermal testing chamber is a light-tight freezer with an internally mounted electric heater. One PMT is tested at a time, and each is housed within an internal dark box. The temperature is controlled using an INKBIRD ITC-308-WIFI temperature controller, which is remotely operated and integrated into an automated DAQ system. 

Four temperature sensors are used, with three placed inside the box next to the PMT bulb. The last temperature sensor is placed above the PMT. The temperature sensors are not placed in direct contact with the photo-cathode. Therefore, there is an expected thermal equilibration time for the cathode to reach thermal equilibrium with the air. This is taken into account by including a 4 hour wait at each temperature, before a run is started. For each voltage measurement, an additional 30 minutes is passed before each measurement to ensure stabilisation of the gain after ramping of the bias voltage. 

\begin{figure}[htb]
    \centering
    \includegraphics[width=0.495\textwidth]{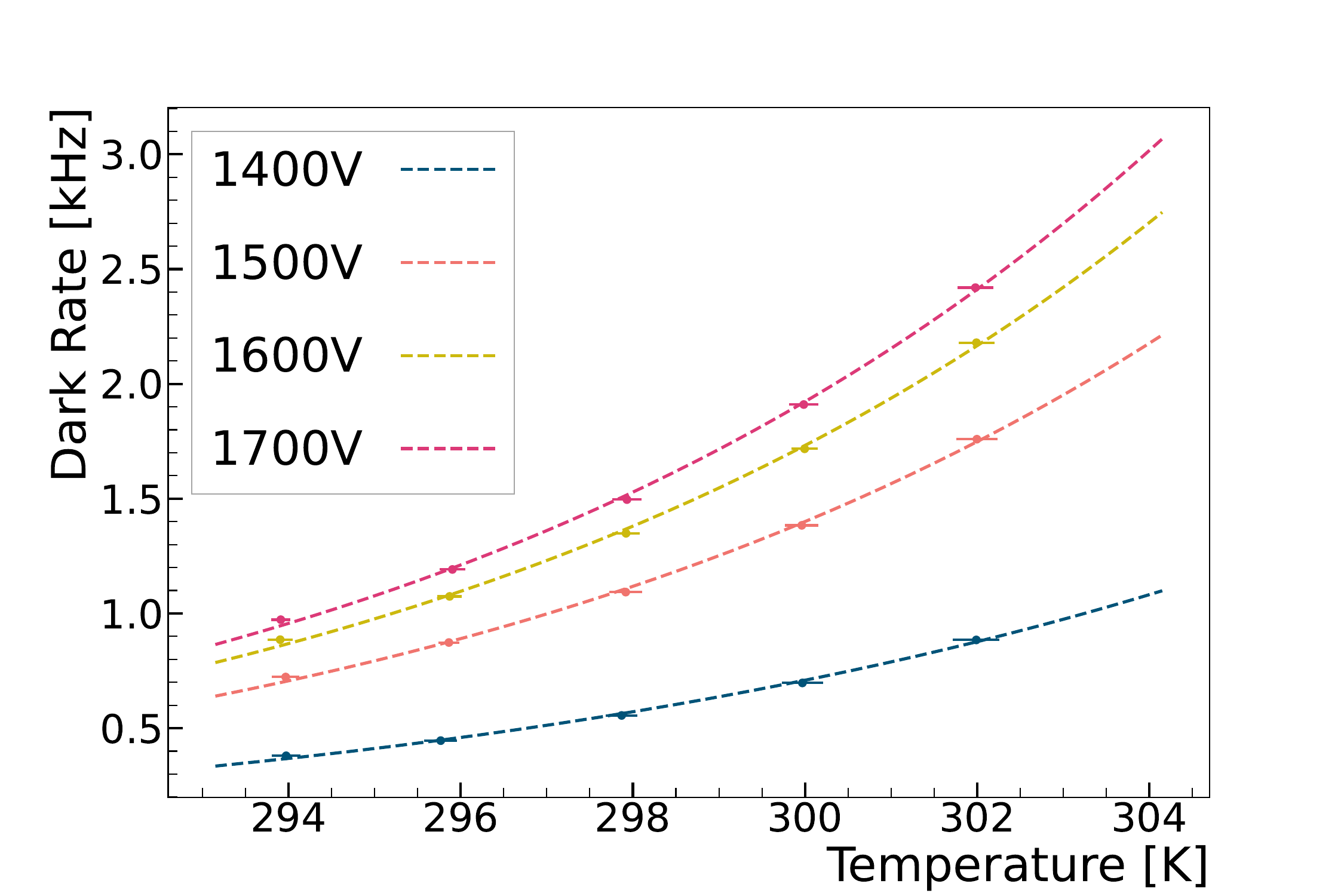}
    \includegraphics[width=0.495\textwidth]{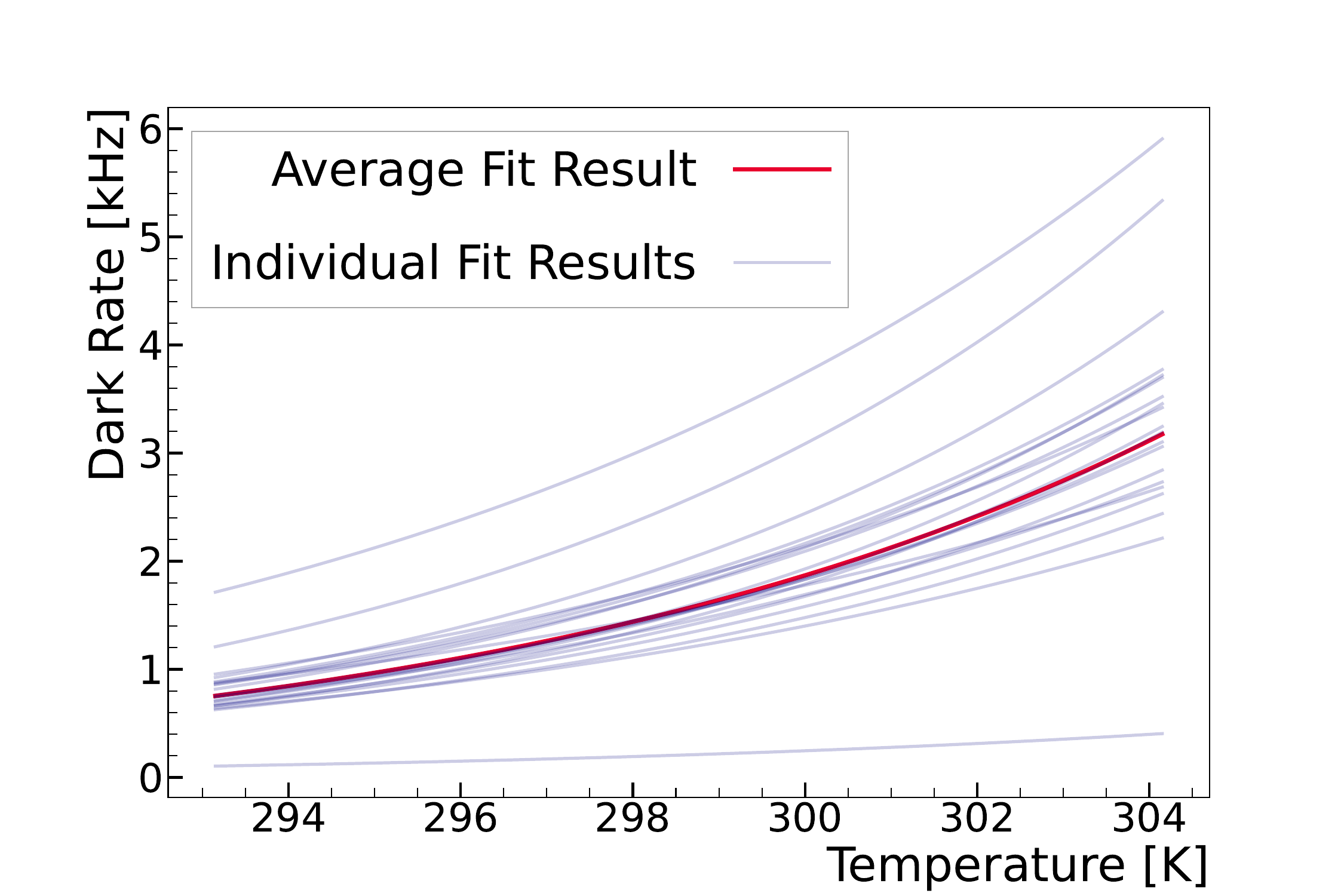}
    \caption{Left: An example of the temperature dependent dark rate fit for the 4 bias voltages and for the listed temperatures, excluding 288.15~K. Right: The average temperature dependent dark rate curve, for 1500~V, alongside with the curves from all 20 PMTs at the same voltage.}
    \label{fig:DRFits}
\end{figure}

\begin{figure}[htb]
    \centering
    \includegraphics[width=0.7\textwidth]{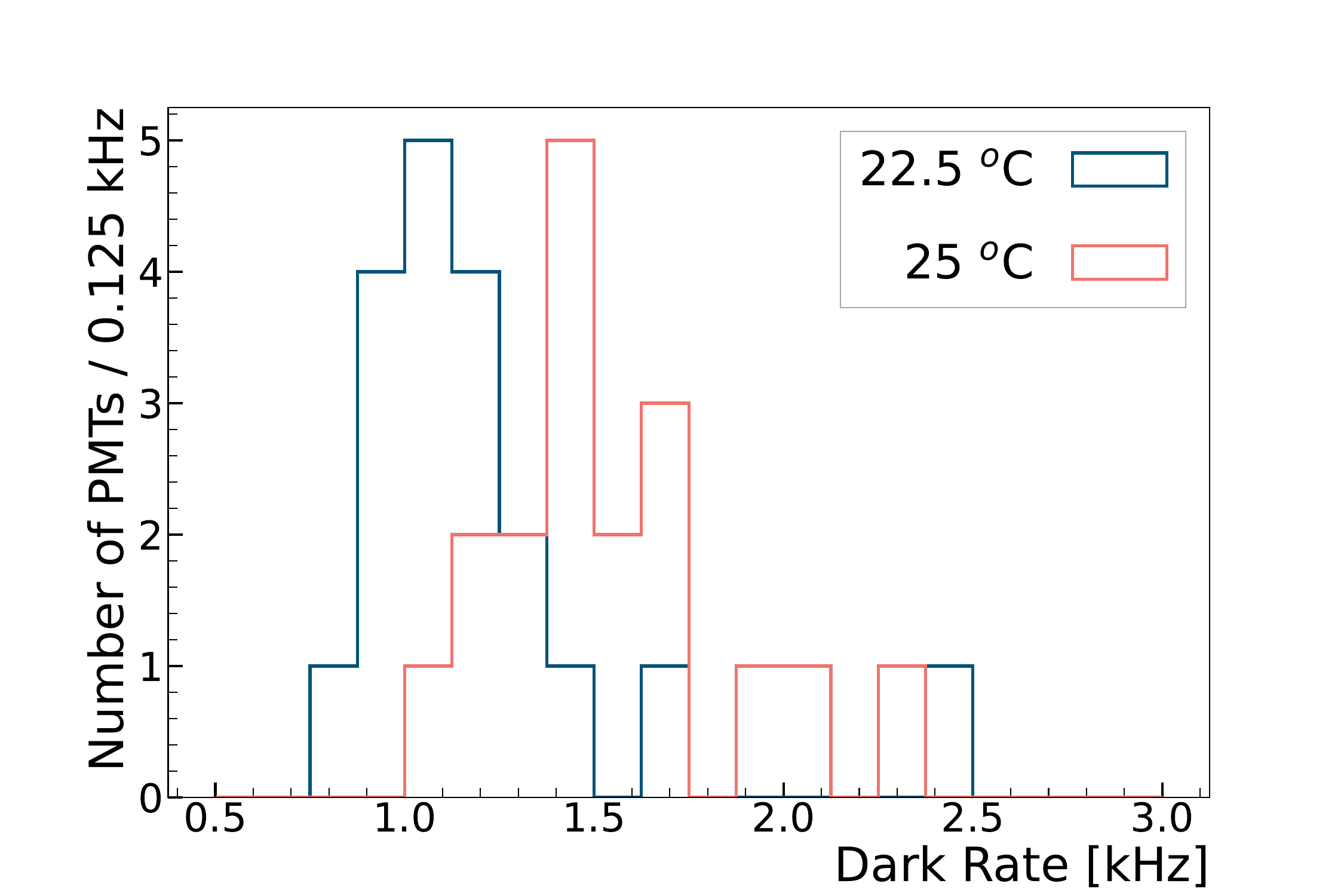}
    \caption{Dark rates for all of the PMTs, at 1500~V and two nominal temperatures --- 295.65~K and 298.15~K.}
    \label{fig:DRRes}
\end{figure}

The results, including the fits of a representative PMT as well as the resulting curves of all 20 (and their average), are shown in Fig.~\ref{fig:DRFits}, where the temperature measurements for each gain value are fitted according to the relationship between dark current and temperature, provided by Hamamatsu. This relationship is defined as:
\begin{equation}
    R = A T^{5/4} e^{-e \psi / k_B T} ,
\end{equation}
where $\psi$ is the work function of the cathode, $e$ the electron charge, and $A$ a constant scale parameter. In the fit, $A$ is floated, along with the parameter $B=-e\psi/k_B$. Statistical errors are evaluated for each rate measurement and are found to be small. Systematic errors associated with the temperature measurement are also evaluated, and are defined as the sum in quadrature of the resolution of the temperature sensor (0.05~K) and half of the range of measurements given by the set of four RTD sensors. The temperature points used for each fit are averages of the time-based averages for each of the three sensors placed within the PMTs box. The dark rate characteristics for each PMT, at the nominal bias voltage of 1500 V for 295.65~K and 298.15~K, are shown in Fig.~\ref{fig:DRRes}.

With these measurements, the expected dark rates can be estimated during runs with the final detector and any associated time-varying contribution to total experimental background. 

\subsection{Timing}

\begin{figure}[htb]
    \centering
    \includegraphics[width=0.495\textwidth]{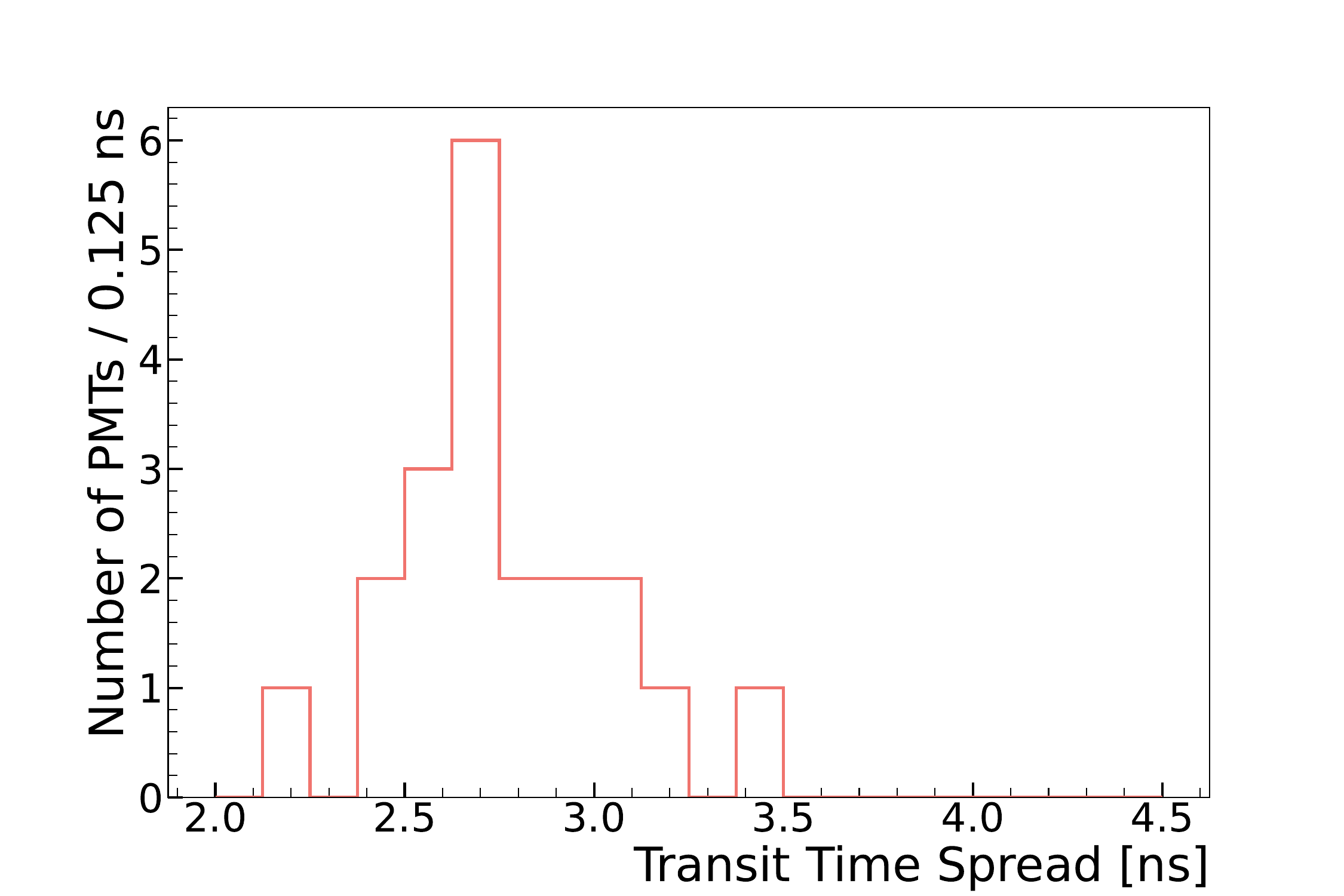}
    \includegraphics[width=0.495\textwidth]{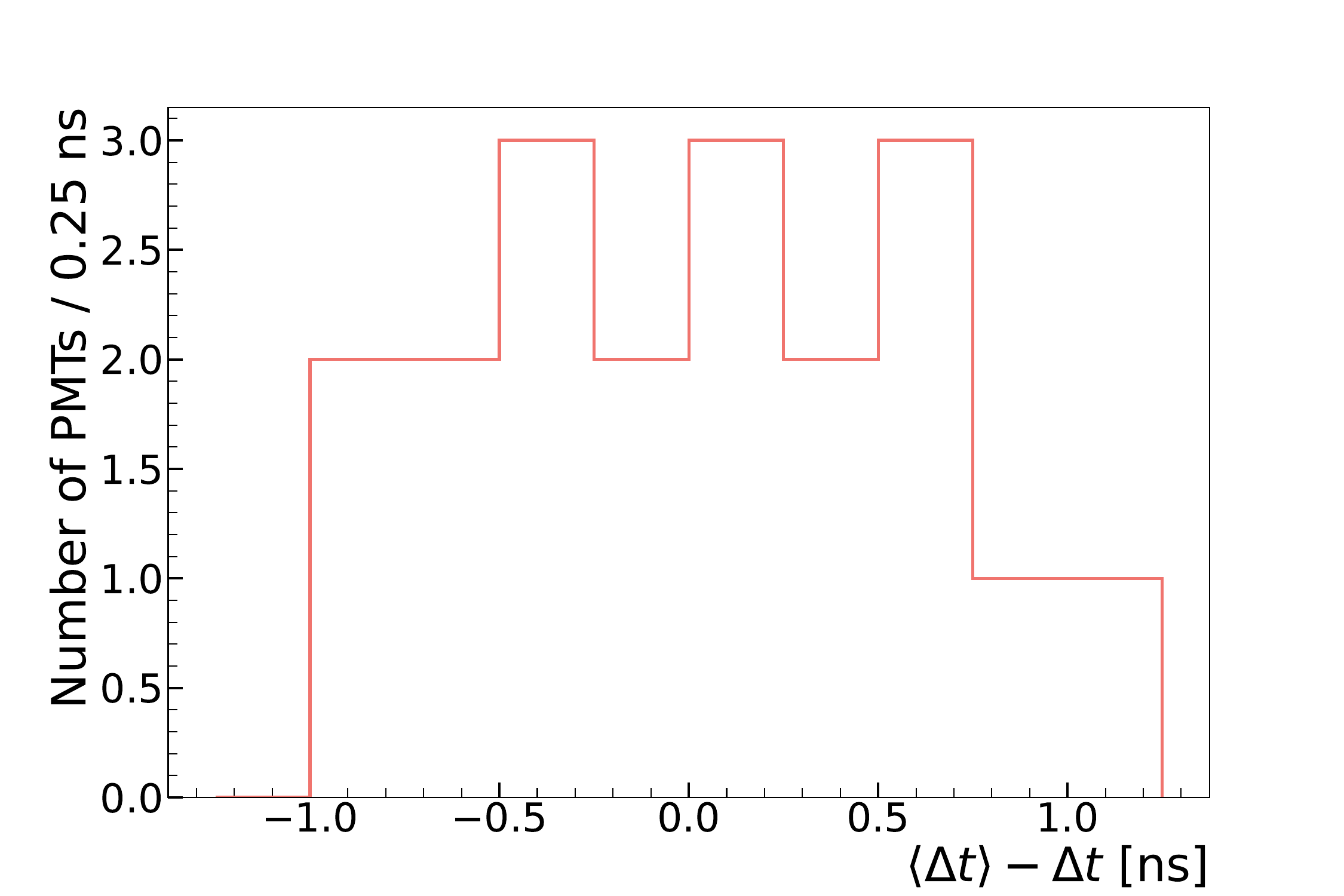}
    \includegraphics[width=0.495\textwidth]{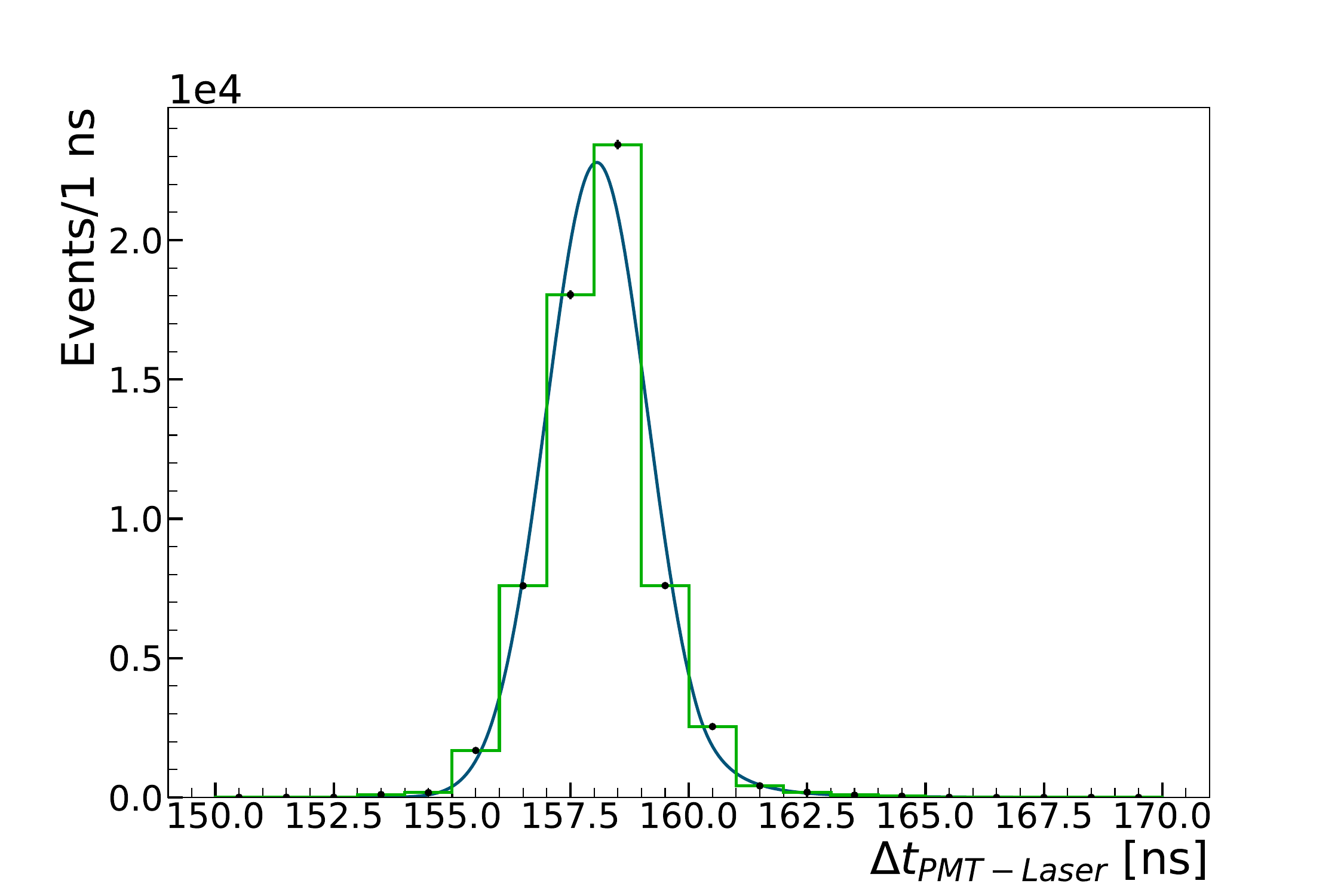}
    \caption{Top Left: A histogram of the TTS values for all 20 of the PMTs, set at the same bias voltage of 1500~V. Top Right: A histogram of the differences between each PMTs relative transit time from the average of the whole set. Bottom: An example fit to a time difference spectrum from one of the R5912 PMTs.}
    \label{fig:HistTiming}
\end{figure}

Knowledge of the timing properties of the PMTs is essential for understanding the position reconstruction capabilities within the veto sub-detector.
With the SPE data-set, the relative transit time amongst the PMTs, and their transit time spread (TTS) can be measured. These are determined from fits to the difference in trigger time between the laser pulse and the corresponding PMT pulse. An example of such a fit is shown in Fig.~\ref{fig:HistTiming}, where the mean and the full width half-maximum correspond to the relative transit time and TTS, respectively. In performing the fit, we bin the time difference spectra in widths of 1~ns, and use zfit~\cite{zfit} to fit the spectra with a Crystal Ball function. While the V1730 digitiser samples every 2~ns, the CFD algorithm used to find the time of each pulse uses a linear interpolation between samples. 

The results of the timing measurement are presented in Fig.~\ref{fig:HistTiming}. These are compatible with the nominal TTS of 2.4~ns reported by Hamamatsu. The TTS values of all PMTs are within 1~ns of the nominal value, with a spread that favours higher values. This is attributed to the trigger pulse of the laser, which itself will have some spread over time that is folded into the TTS. The width of the laser trigger pulse is quoted to be $5\pm3$~ns~\cite{PLP_10Laser}. The variation in transit times is evaluated by subtracting each time-difference value from the average. The transit times have a maximum variation from the average of approximately 1~ns.

\subsection{Detection Efficiency}

The detection efficiency (DE) of the R5912 PMTs is measured relative to the calibrated reference R11065 PMTs from Hamamatsu. To perform a relative measurement, a beam splitter is used to transmit signals to both the reference R11065 PMT and the R5912 PMT simultaneously. To control for the possibility of uneven light splitting, the R5912 and R11065 are both mounted in the position parallel to the laser beam, such that the counts from two separate measurements are compared. An additional R11065 reference is placed perpendicular to the beam, in the other position relative to the beam splitter, to ensure the stability of the light source between runs. We refer to a DE in this case, rather than a quantum efficiency (QE), as the collection efficiency of the first dynode for the R5912 PMT is not known. The collection efficiency of the first dynode refers to the percentage of PEs that reach the dynode and undergo amplification. Hamamatsu generally guarantees values between 80-100\%, but this may vary depending on the design of the dynode structure. The R5912 PMT uses a box and linear-focused design, which Hamamatsu describes as having a ``good'' collection efficiency~\cite{HamamatsuHandbook}. The design that maximises collection efficiency is the box and grid design, which is what the R11065 PMTs use. Thus, we reasonably approximate the DE and QE of the R11065 to be equal.

To reduce the impact of background noise in the measurement, only signals that correlate in time with the laser signal are considered --- this selection is performed in the same way as in Sec.~\ref{sec:SPE}. As some additional low-amplitude noise remains in the both the R11065 and R5912 samples, some additional peak height and charge cuts are applied. The charge cuts require that all events have charge values above 0, whilst a peak height cut of $A>2.5$~mV was applied to both PMTs. Finally, the datasets were time normalised by requiring that the timestamps for the final events in each dataset are approximately equal. The remaining events are then counted, and input into the below equation to determine the relative DE of the R5912 PMT:
\begin{equation}
    {\rm DE}_{R5912} = \frac{N_{R5912}}{N_{R11065}} \times {\rm QE}_{R11065}
\end{equation}
where ${\rm QE}_{R11065} = 0.3325$. 

Using this method, we measure that ${\rm DE}_{R5912} = 0.2862 \pm 0.0009$, where the error is statistical, and the measurement is performed for 405~nm. This is calculated from $N_{R5912} = 196746$ and $N_{R11065} = 228507$. The nominal QE given by Hamamatsu, which is for 390~nm, is 0.25~\cite{veto-pmts}, which is comparable to the measurement above. To attain a broader understanding of the DEs of the remaining R5912 PMTs, we use the above PMT (for which the relative DE above was determined) as a reference. Given there are common SPE datasets, we make the comparison using that data, but only among the PMTs which shared the same position as the reference in PMT in the setup shown in Fig.~\ref{fig:SPESetup}. As a result, we can find the relative DEs for 9 PMTs. To make the comparison, we find the fractional yield of the SPE gaussian fitted to the charge spectrum (fractional with respect to the total number of events in the charge spectrum), and compare it to the fractional yield from the reference PMT. The resulting number is the relative DE of the PMT. Due to the threshold artefact discussed in Sec.~\ref{sec:SPE}, however, there is a variation in the determined relative DE depending on the chosen bias voltage/gain. This is a systematic error, which we quantify by taking half of the range in which the relative DE values varied. Both the values for the relative DE, and the systematic error, are shown in Table~\ref{tab:rel-qes}. There is good consistency seen amongst the DEs for each PMT. 

\begin{table}[htb]
\centering
\caption{The relative DE for each of the 9 PMTs also placed in the same position as the reference (according to Fig.~\ref{fig:SPESetup}), during the acquisition of the SPE data analysed in Sec.~\ref{sec:SPE}. The systematic errors are also included. The chosen bias voltage for which these relative QEs are determined is 1700~V.}
\begin{tabular}{l|ll}
PMT Serial Number                  & Rel. DE & Sys. Err.\\ \hline
KQ0059                                              & 1.11 & 0.045  \\
KQ0160                                  & 0.95 & 0.038 \\
KQ0143                                                 & 1.00 & 0.020 \\
KQ0159
         & 1.01 & 0.059  \\
KQ0145
         & 1.02 & 0.037  \\
KQ0147                                     & 1.01 & 0.019 \\
KQ0148                                          & 1.00 & 0.018 \\
KQ0163                                          & 0.96 & 0.038  \\ 
KQ0157                                          & 1.04 & 0.032  \\ 
\hline
\end{tabular}
\label{tab:rel-qes}
\end{table}

\subsection{Linearity and Saturation}
\label{sec:LinandSat}

The dynamic range and saturation for a subset of the R5912 PMTs were measured in a setup similar to that used for the SPE measurements, but with the beam-splitter removed in order to maximise the intensity of the beam that reaches the PMT, and an adjustable aperture in place of the fixed diameter pin-hole. Differing levels of intensity were achieved by varying the attenuation of our optical filters and the power of the laser source. The aperture was kept at a fixed diameter that ensured both high enough intensities and a small enough beam size/spread such that the entire beam passes through each filter. To ensure that each measurement (with each of the optical filters) is performed in the same stable dark box environment, each filter was mounted sequentially on a Thorlabs LTS150/M linear stage, programmed to step through each mounted filter so that data can be collected for each without manually disturbing the dark box environment. The optical densities (OD) for each of the filters were: 0.5, 1.0, 1.5, 2.0, 2.5. Similar methods have been used in previous studies~\cite{ProtoDUNELin}, and were chosen for this measurement due to being restricted to one light source.

For each point corresponding to each filter, an expected intensity is calculated based on applying the OD of the filter to
a PE count (determined from the SPE response) for a low attenuation point, which can be expressed as
\begin{equation}
    I_{{\rm exp}, x} = \frac{q_{\rm ref}}{Q_{\rm SPE}} \times 10^{\rm{OD}_x},
\end{equation}
where $x$ corresponds to the given ND filter, $q_{\rm ref}$ is the charge of the low intensity reference point, $Q_{\rm SPE}$ is the mean SPE charge, and $\rm OD$ is the optical density of the given filter. This is compared to a measured intensity, which is defined as the photon count derived by directly from the SPE response for the corresponding PMT, expressed as
\begin{equation}
    I_{\rm meas, x} = \frac{q_{x}}{Q_{\rm SPE}},
\end{equation}
where $q_x$ is the charge measured using a given ND filter.

\begin{figure}[htb]
    \centering
    \includegraphics[width=0.7\textwidth]{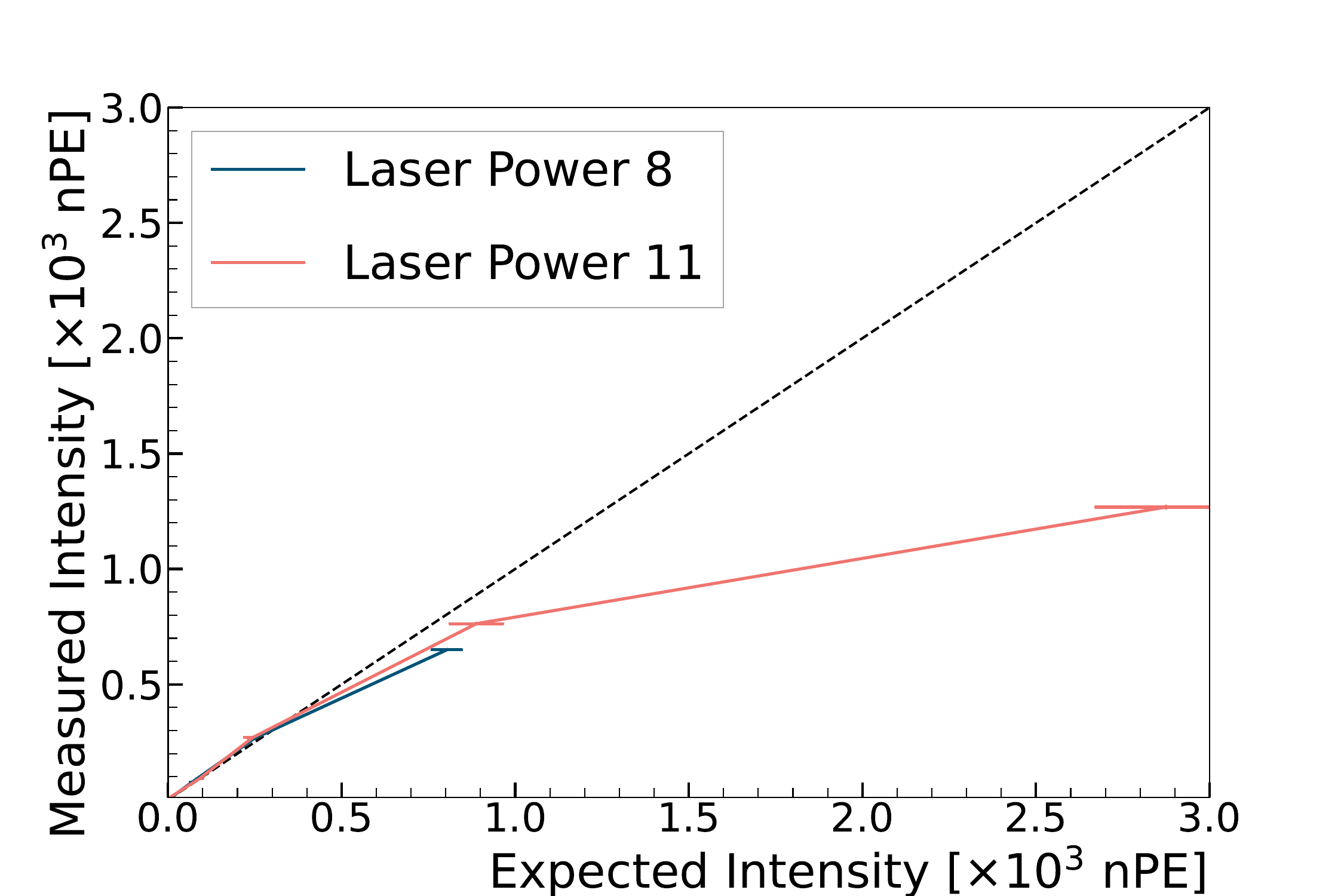}
    \caption{Measured and expected intensity plotted against each other for each dataset, and for two laser powers, to visualise the response curve of the PMT. The charge response of the PMT begins to saturate at $\approx$500~PEs, and is linear below this charge.}
    \label{fig:LinResponse}
\end{figure}

\begin{figure}[htb]
    \centering
    \includegraphics[width=0.7\textwidth]{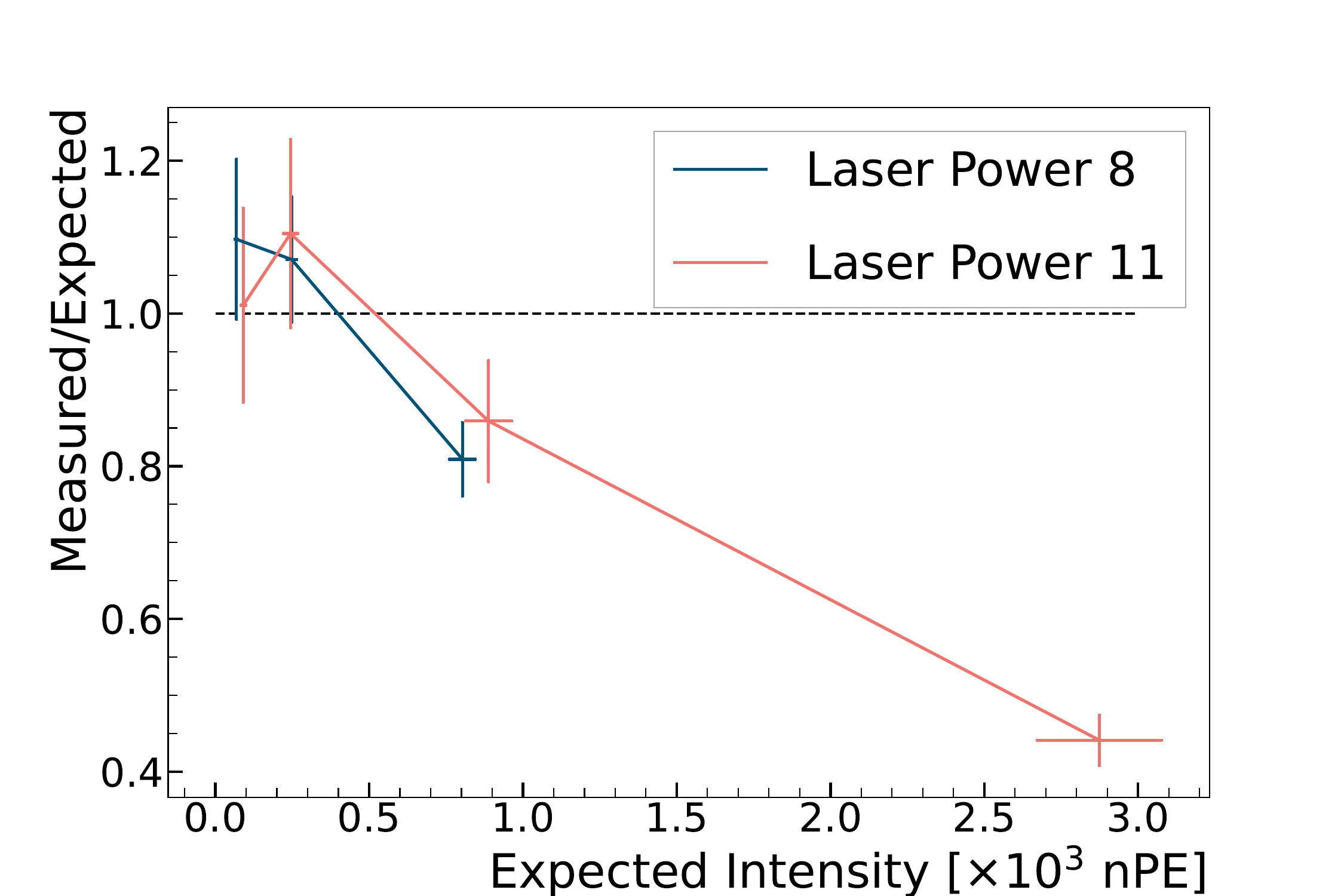}
    \caption{The ratio of the measured and expected intensities, plotted against the expected intensity, for each ND filter dataset and for two laser powers. A ratio compatible with 1 corresponds to the linear response regime, with lower values caused by the non-linear response/saturation of the PMT. The cause for values above 1 is due to the systematic error arising from the tolerance of the OD values for each filter, which is included in the error bars.}
    \label{fig:LinRatio}
\end{figure}

Each intensity can be plotted against each other to demonstrate the linear and non-linear region of the PMT responses, shown in Fig.~\ref{fig:LinResponse}. However, Fig.~\ref{fig:LinRatio} shows that the ratio of the two can also be taken to quantify the degree of non-linearity once the PMT response exits the linear regime. 

The intensity of background signals within the SABRE South veto system is likely to vary greatly --- from high energy muons at the GeV-scale, to lower energy gamma radiation from crystal impurities. On average, high MeV to GeV-scale signals will be required for saturation to be seen, based on the average number of detected PEs per keV. Proximity to sensors may also impact this. As a result, it is important to understand at what photon intensity will each sensor saturate, and then to correct it during analysis, such that the energy scale of signals can be properly understood. Correction factors can be determined from the results shown in Fig.~\ref{fig:LinRatio}, for example. 

\subsection{Afterpulsing}

Afterpulses typically follow a high intensity initial pulse, with a consistent and well-defined time gap between the afterpulse and main pulse. Understanding the contribution of these afterpulses to the total detector noise, in the form of an afterpulsing probability per PE, is important for both background modelling of the veto system and also modelling of the PMT performance for digitisation studies. These afterpulses can become a form of background when the time difference from the main pulse is large enough to span multiple event/waveform windows. The mechanism producing these afterpulses is driven by small amounts of gas molecules in the vacuum of the PMT, which are ionised by the current induced by the initial signal. The ions drift under the electric field of the PMT and generate their own pulses upon collision with the cathode or dynode, the species of the ion, and the electric field structure of the PMT is what determines the time difference between the main pulse and said afterpulse.

We measure the afterpulsing in two of the PMTs, reusing the setup used to measure the response linearity (see Sec.~\ref{sec:LinandSat}). This setup allows for measuring the fraction of events within a run that have at least one afterpulse (afterpulsing fraction) for a range of different initial pulse intensities. From these data, an afterpulsing probability per PE can be determined. The waveform window is set to 8192 samples, or 16.384~$\mu$s, to ensure any possible afterpulses are caught in the same event window. Afterpulses are identified using the SciPy peak finder algorithm~\cite{scipy}, and are only considered if they occur at least 500~ns after the initial signal peak, have an amplitude greater than 1.5~mV, and a peak prominence of greater than 15. Once any afterpulses have been identified, their time separation from the main pulse is determined.

\begin{figure}[htb]
    \centering
    \includegraphics[width=0.54\textwidth]{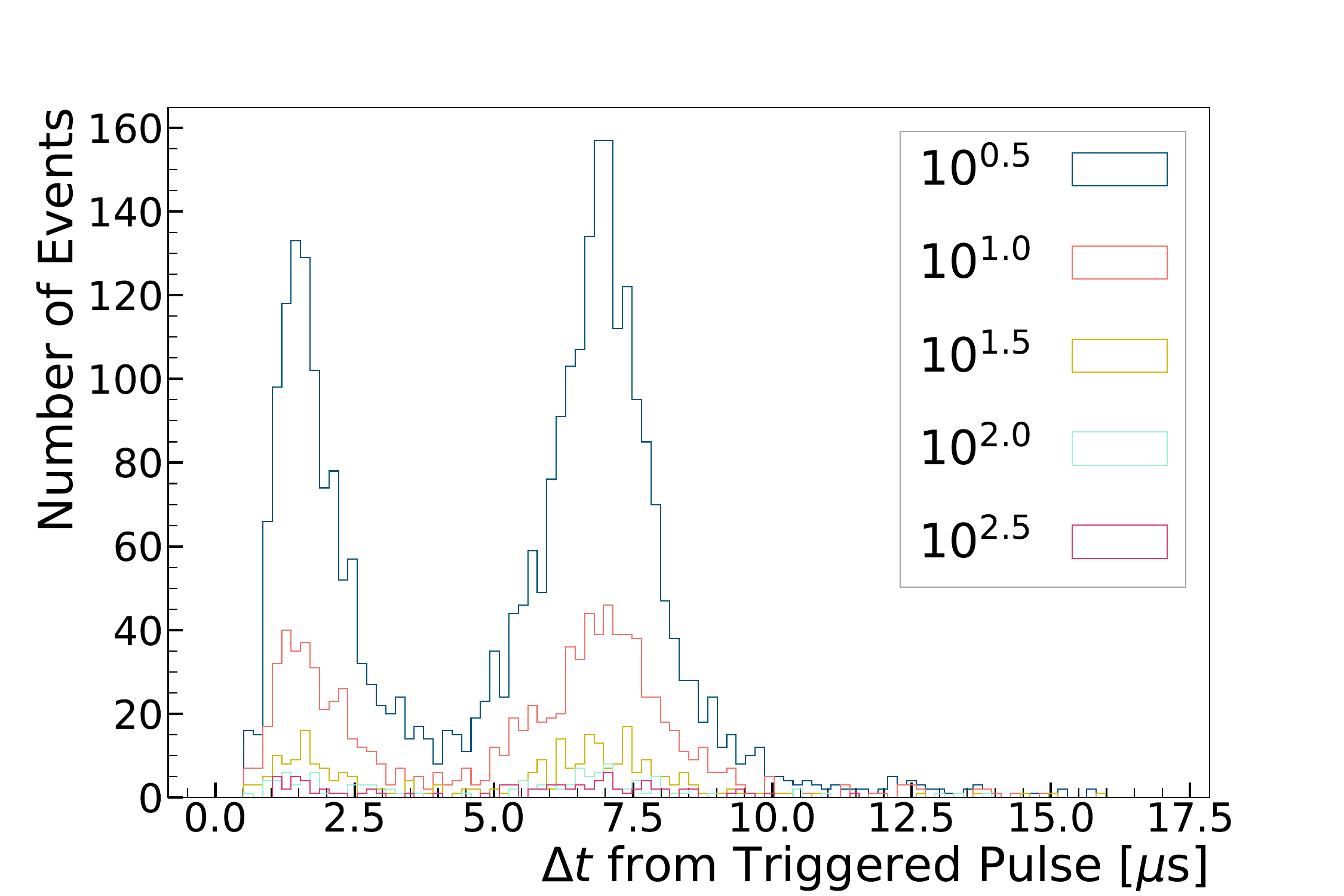}
    \includegraphics[width=0.45\textwidth]{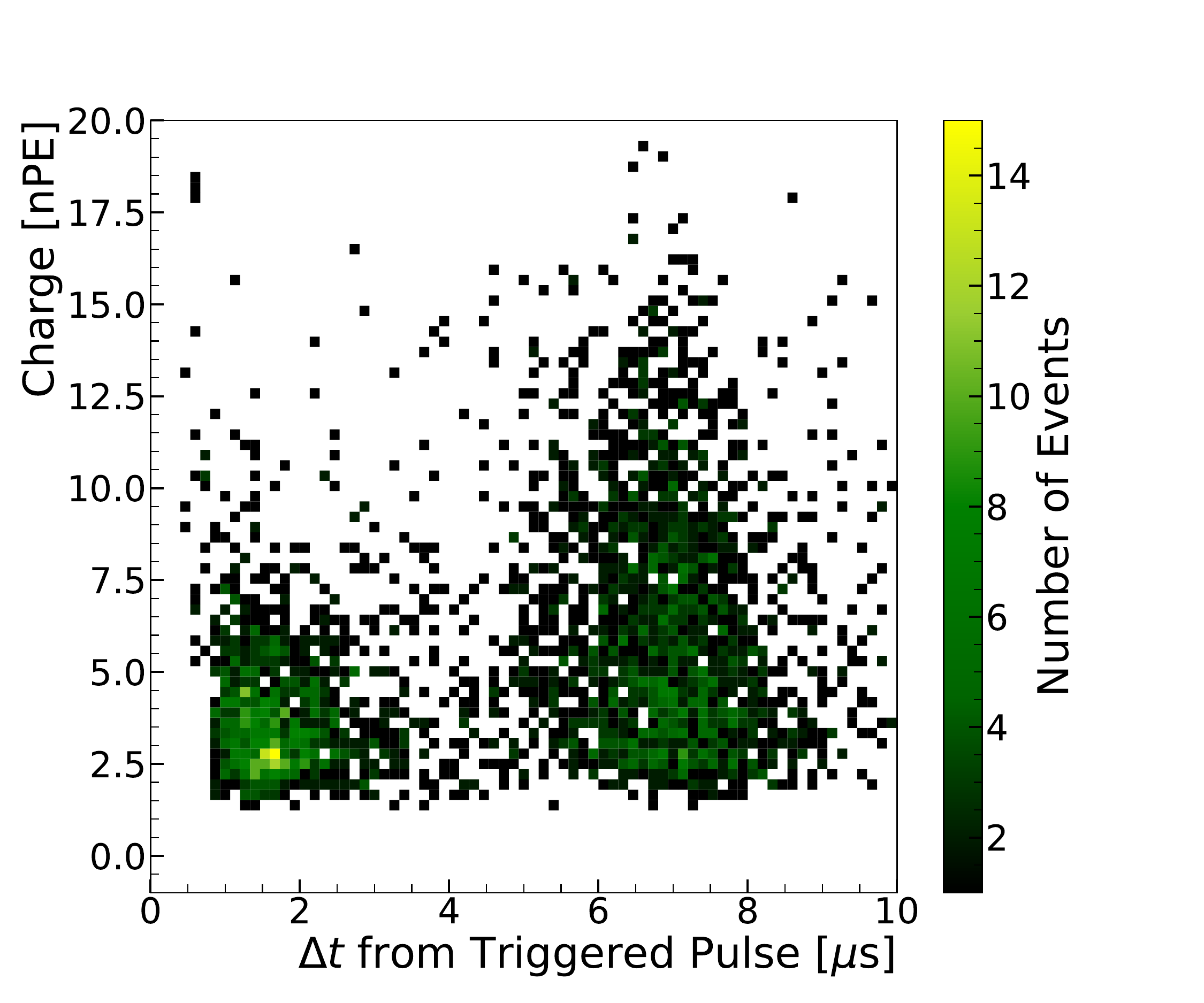}
    \caption{Left: The afterpulse time difference spectra for one the SABRE R5912 PMTs. Two populations can be noticed, with the majority of afterpulses occurring after 2~$\mu$s. These can potentially be attributed to He or CH$_4$ ions. The legend refers to the attenuation factors of the five ND filters used. Right: Time difference from triggered pulse vs. pulse charge in units of PE, using the dataset with lowest (0.5) OD filter. Notably, majority of pulses are above the charge of a SPE.}
    \label{fig:APulseDiffs}
\end{figure}

Two key populations of time differences are found Fig.~\ref{fig:APulseDiffs}, both of which are outside the maximum nominal waveform window for SABRE South (1024 samples, or 2.048~$\mu$s). Combining this with charge, Fig.~\ref{fig:APulseDiffs} also shows that all afterpulses occur with charges that are within 1-11 PE, and so can be easily misidentifed as real signals --- particularly if they are over 1 PE, so are unlikely to be considered dark rate. 

\begin{figure}[htb]
    \centering
    \includegraphics[width=0.7\textwidth]{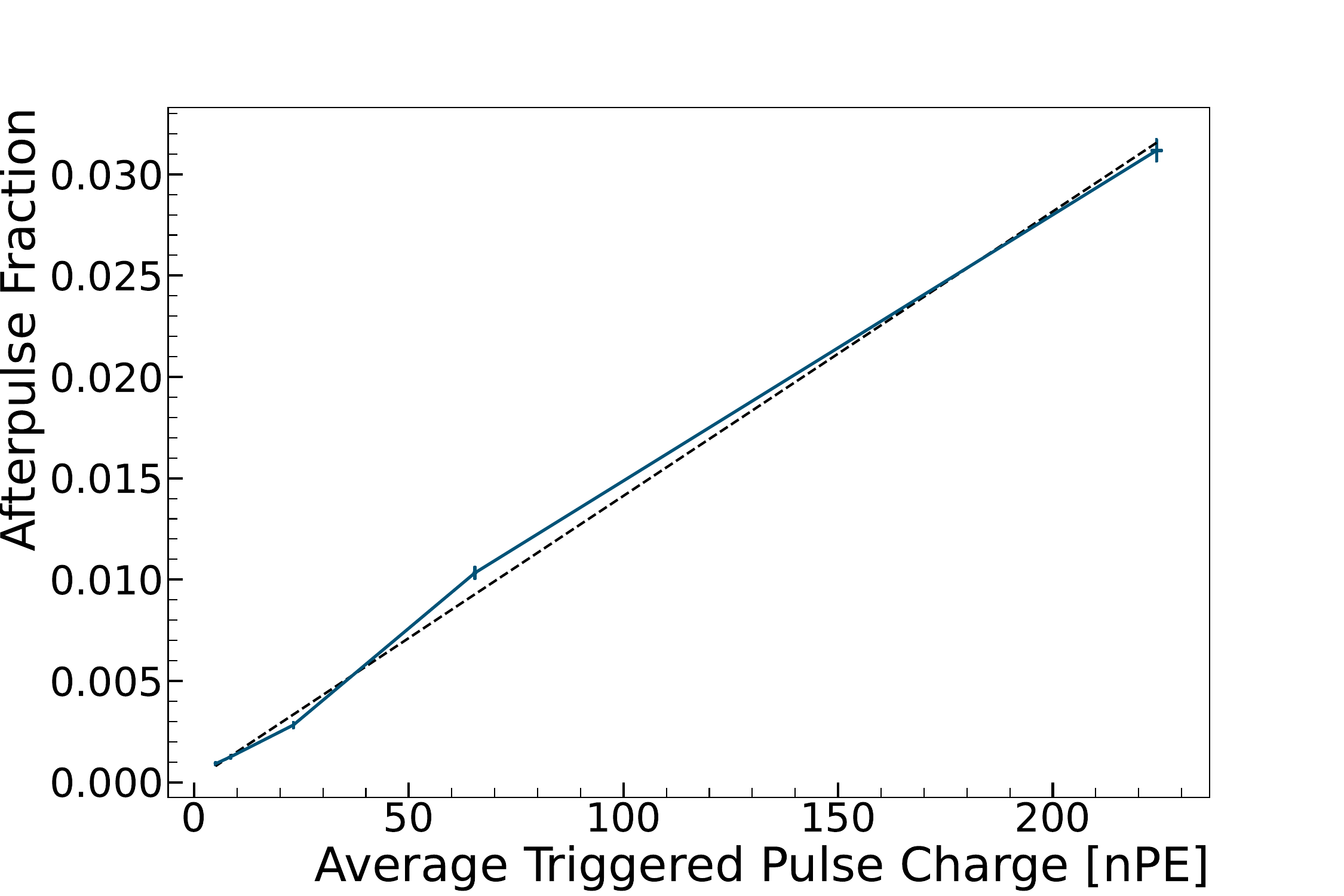}
    \caption{Fraction of events with an afterpulse for each given ND filter dataset, as a function of the average triggered pulse charge for the dataset corresponding to each filter. The line of best fit is overlaid.}
    \label{fig:APulseProb}
\end{figure}

To determine the probability of afterpulsing per PE, we plot the afterpulsing fraction as a function of the initial pulse charge, perform a linear regression using the lmfit package~\cite{lmfit} and extract the resulting gradient of the fitted line as the parameter of interest. The linear expression used for the fit is
\begin{equation}
    f_{\rm AP} = p_{\rm AP}q + c,
\end{equation}
where $f_{AP}$ is the fraction of events containing an afterpulse, $q$ is the charge of the triggered pulse, $p_{\rm AP}$ is the afterpulsing probability in units of $\rm PE^{-1}$, and $c$ is an arbitrary axis intercept that should be close to 0. The result of this fit is shown in Fig.~\ref{fig:APulseProb}, with the resultant probability being $p_{\rm AP} = (1.7 \pm 0.08) \times 10^{-4}$~afterpulses/PE (with the uncertainty being the fit error). The full expression of the linear fit can be written as $f_{AP} = (1.7q + 1.7)\times 10^{-4}$. This parameter helps us measure the inherent noise of the PMT for digitisation of simulated PMT signals, compare between PMTs, and to understand the level of potential contamination during storage and production. It can also act as a benchmark for subsequent characterisation with the veto's optical calibration system, which allows us to monitor ingress of gas into the PMT during the lifetime of the experiment.

\subsection{Spontaneous Light Emission in Oil-proof Base}

Previous liquid scintillator-based experiments have reported spontaneous light emission from their PMTs in-situ, an effect induced by the electronics present in the oil-proof bases. The Double Chooz experiment has characterised this effect \cite{DoubleChooz:2016ibm}, after finding a coincident background rate associated with the number of PMTs powered at any time, along with their bias voltage. A study of this effect yielded an exponential dependence on bias voltage and temperature. Further investigation allowed the Double Chooz experiment to determine that the effect originates from coronal light discharges due to the polarisation of epoxy under the presence of circuitry in the base, causing air in the gaps between the epoxy and circuit to glow under the presence of a strong field \cite{DoubleChooz:2016ibm}. The SABRE South R5912 PMTs make use of the same epoxy compound, and although air gaps between the circuitry and epoxy were minimised during production to mitigate this effect, a study to test for a reduced rate is still required.

To detect the light emission from the base, we placed one of the R5912 oilproof PMTs in a dark box, with the base end facing a Hamamatsu R11065 PMT. These measurements were performed at room temperature. 
With a threshold of 3.7 mV, data was initially collected with the R5912 not powered, allowing a determination of the dark rate at 23.85 Hz. Biasing the R5912 PMT caused the rate in the observing PMT to increase. This test was performed at a range of voltages, to test the dependence of light emission on the bias voltage. 

\begin{figure}[htb]
    \centering
    \includegraphics[width=0.59\textwidth]{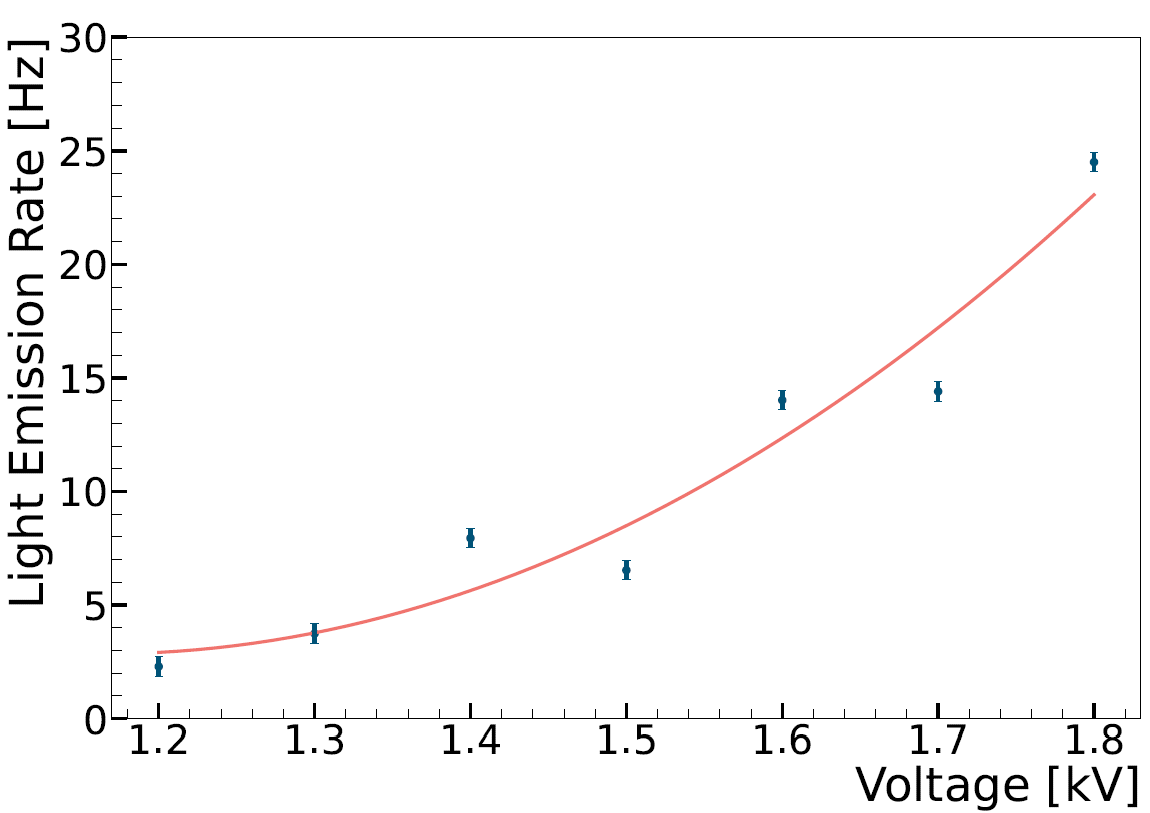}
    \caption{Rates attributed to spontaneous light emission in the R5912 oilproof base, at a range of different voltages, showing an increasing relationship between rate of light emission and bias voltage. An exponential curve is loosely fitted.}
    \label{fig:light_emission_Vcurve}
\end{figure}

A rate attributable to spontaneous light emission in the base was found, with an observed dependence on bias voltage, shown in Fig. \ref{fig:light_emission_Vcurve}. Measures to mitigate this effect will be implemented in the final detector, such as light-proofing.

\section{Particle Discrimination in Liquid Scintillator Prototype}

To test the performance of an R5912 PMT in a situation analogous to the final detector, a liquid scintillator detector was constructed with direct coupling of the PMT with a liquid scintillator mixture similar to that used for the final experiment. This test detector was used to study the efficacy of neutron/gamma discrimination methods using the R5912 PMTs. Neutrons were chosen as a particle of interest because neutrons can mimic a DM signal and be present within the detector due to neutron spallation from cosmogenic muons~\cite{neutronspall}. Gamma rays are of interest due to their association with radioisotope decays that include products with energies within the 1--6~keV$_{\rm ee}$ signal region of interest --- for example, the $^{40}$K decay which emits an auger electron within 1--6~keV and a gamma ray that can be tagged in the liquid scintillator veto. Using the liquid scintillator veto to understand these backgrounds is advantageous and creates a pathway to linking understood interactions in the veto with correlated events in the crystal detectors. We also discuss methods of removing electronic noise, particularly from a SPE dataset, using variables defined in the frequency domain.

\subsection{Detector Design}

\begin{figure}[htb]
    \centering
    \includegraphics[width=0.4\textwidth]{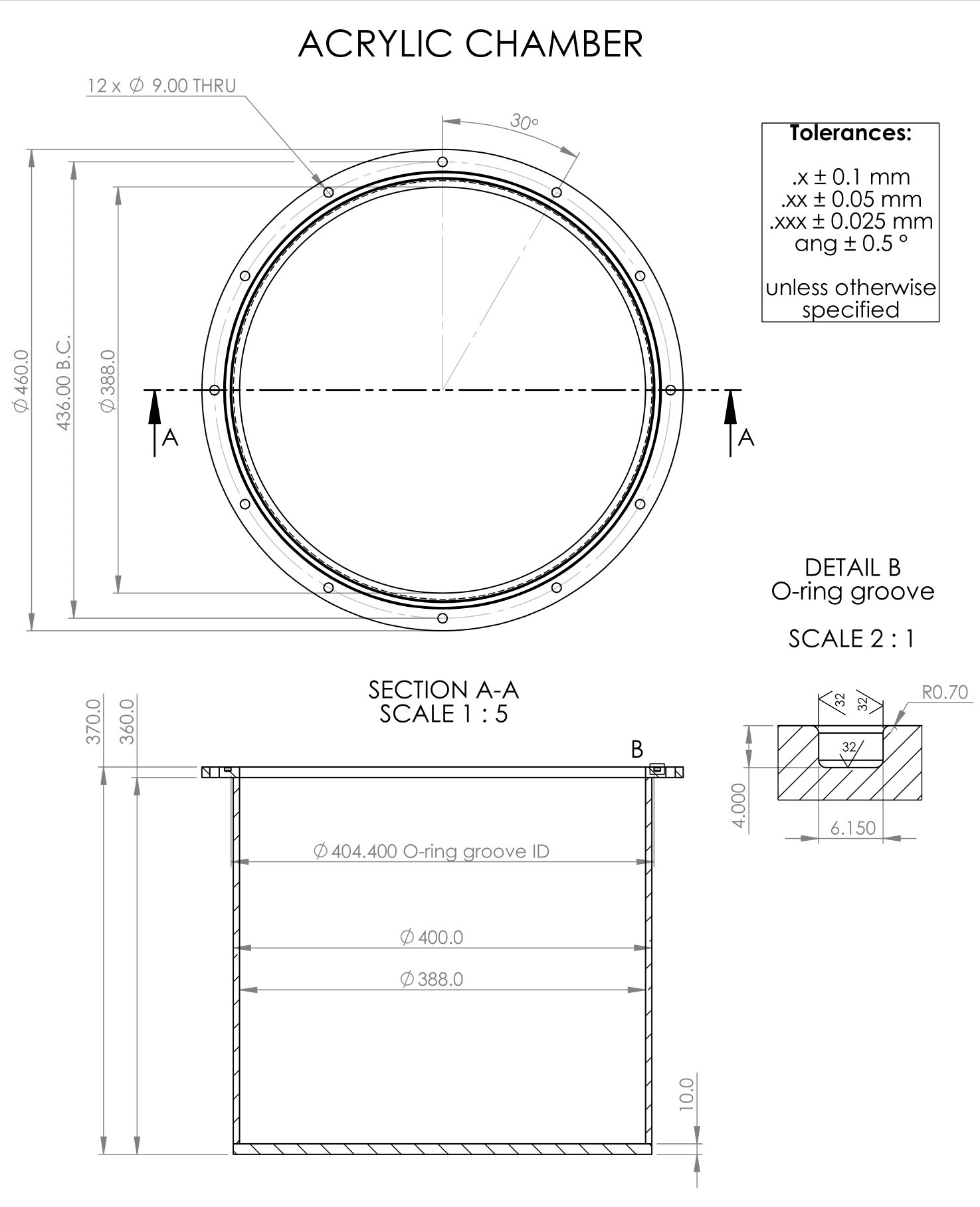}
    \includegraphics[width=0.4\textwidth]{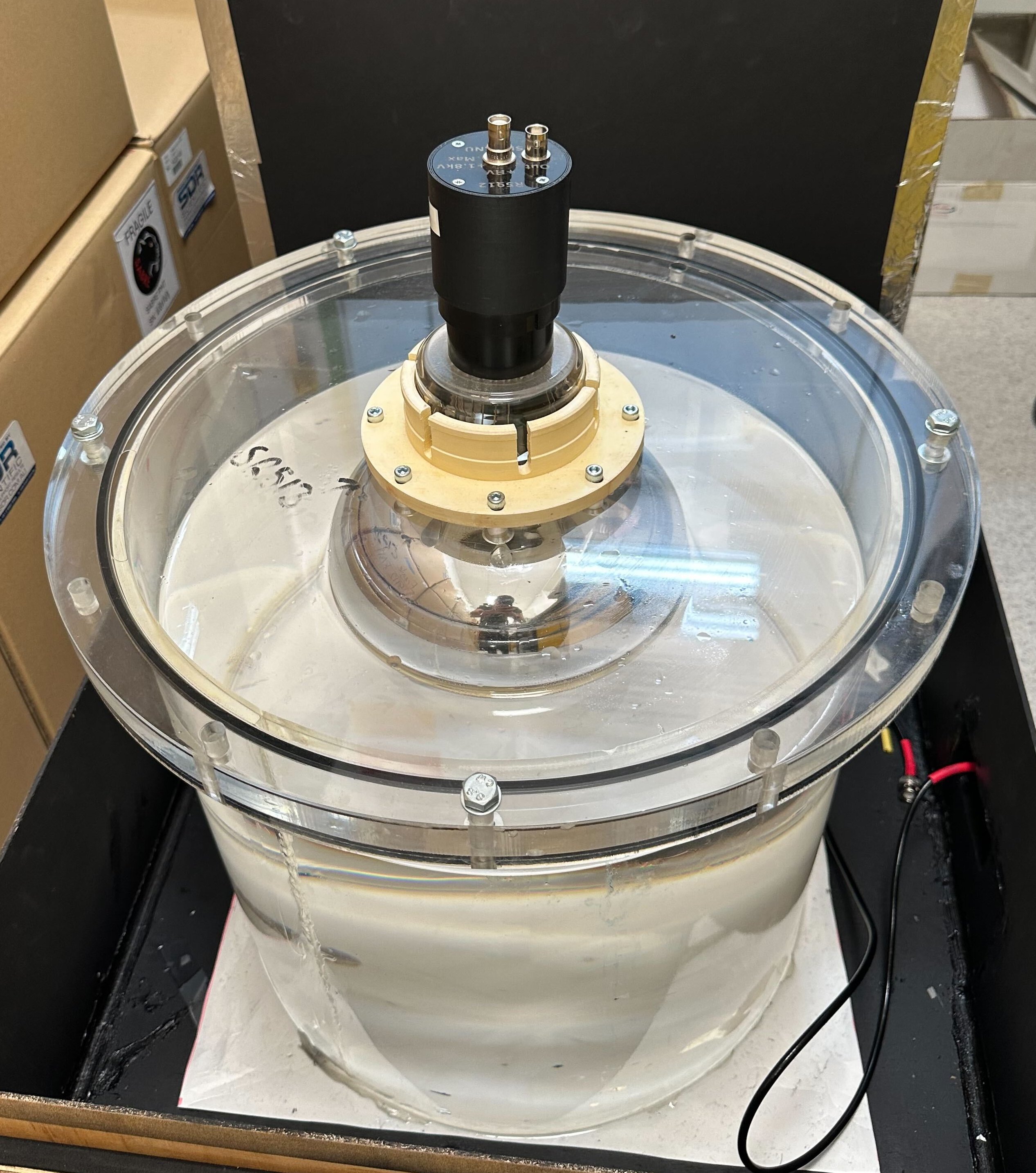}
    \caption{A schematic of the acrylic vessel used to contain the liquid scintillator, alongside a photo without the Lumirror wrapping.}
    \label{fig:SabrSchem}
\end{figure}

The prototype detector (Fig.~\ref{fig:SabrSchem}) is constructed from a cylindrical PMMA acrylic vessel, containing 32.5 L of linear alkylbenzene based liquid scintillator, doped with 3~g/L of PPO and 15~mg/L of Bis-MSB. The PMT is mounted in the top of the vessel via a flange containing an O-ring, with the photocathode directly in contact with the liquid scintillator. Due to the PMT's bouyancy an internal support was not necessary, but was included. The lid is sealed to the vessel via an O-ring, to prevent air leakage and the dissolution of oxygen in the liquid scintillator, affecting the light yield~\cite{Lombardi:2019epz}. Lumirror was wrapped around the main barrel of the vessel and placed above and below the lid and base of the vessel, respectively, to ensure optimal light collection efficiency. The R5912 PMT used did not have an oilproof base but it had the same circuit as the oilproof bases. The PMT was biased to 1300~V for all measurements, and used the same HV system as used for the PMT characterisation. Its mean SPE charge at 1300~V is 0.89~pC.

Waveforms were collected using a CAEN V1730 digitiser connected to a PC via optical link. For time-of-flight studies, a coincidence between a muon detector and the test vessel was required. The analysis was performed using the SABRE South analysis framework, Pyrate. 

\subsection{Tagged Event Selection}

Two setups were used for data collection: a setup for measuring time-of-flight, with a single channel muon detector placed above the prototype detector; and a setup to acquire a gamma ray sample using a radioactive source. 

Time of flight information is used to collect tagged samples of neutrons. An Americium-Beryllium (Am-Be) source was used to obtain neutron data. Neutrons are produced from this source via the capture of an alpha particle (emitted from the Am nucleus) by the Be nucleus. This mechanism also results in the emission of a gamma ray. Placing the Am-Be source above the plastic scintillator based muon detector, the time-of-flight between the two detectors can be determined. The resultant spectrum can be used to separate the neutrons from Am-Be gammas and background muons. Furthermore, the coincidence requirement limits background contamination. The DAQ was configured with a 60~ns coincidence window between the two detectors (each single channel), with a 10~mV threshold in both channels. The prototype detector's PMT was set to a bias voltage of 1300~V, to keep the PMT in its linear range, and the muon detector PMT was set to 1500~V. 
A clean gamma ray dataset was acquired via placing a $^{60}$Co source directly ontop of the lid of the liquid scintillator detector, with the beta particles not able to pass through the acrylic. The prototype detector was triggered at a threshold of 10~mV.

\subsubsection{Selection Criteria}

Additional event selection is performed on each dataset to obtain clean samples of each particle type. In both datasets, the charge is calculated using an event window that starts 20~ns before the start of the pulse (identified via a interpolated leading edge threshold algorithm), and ends 486~ns after the pulse start. The leading edge threshold for the $^{60}$Co dataset is 25~mV, and is 50~mV for the Am-Be dataset.

\begin{figure}[htb]
    \centering
    \includegraphics[width=0.495\textwidth]{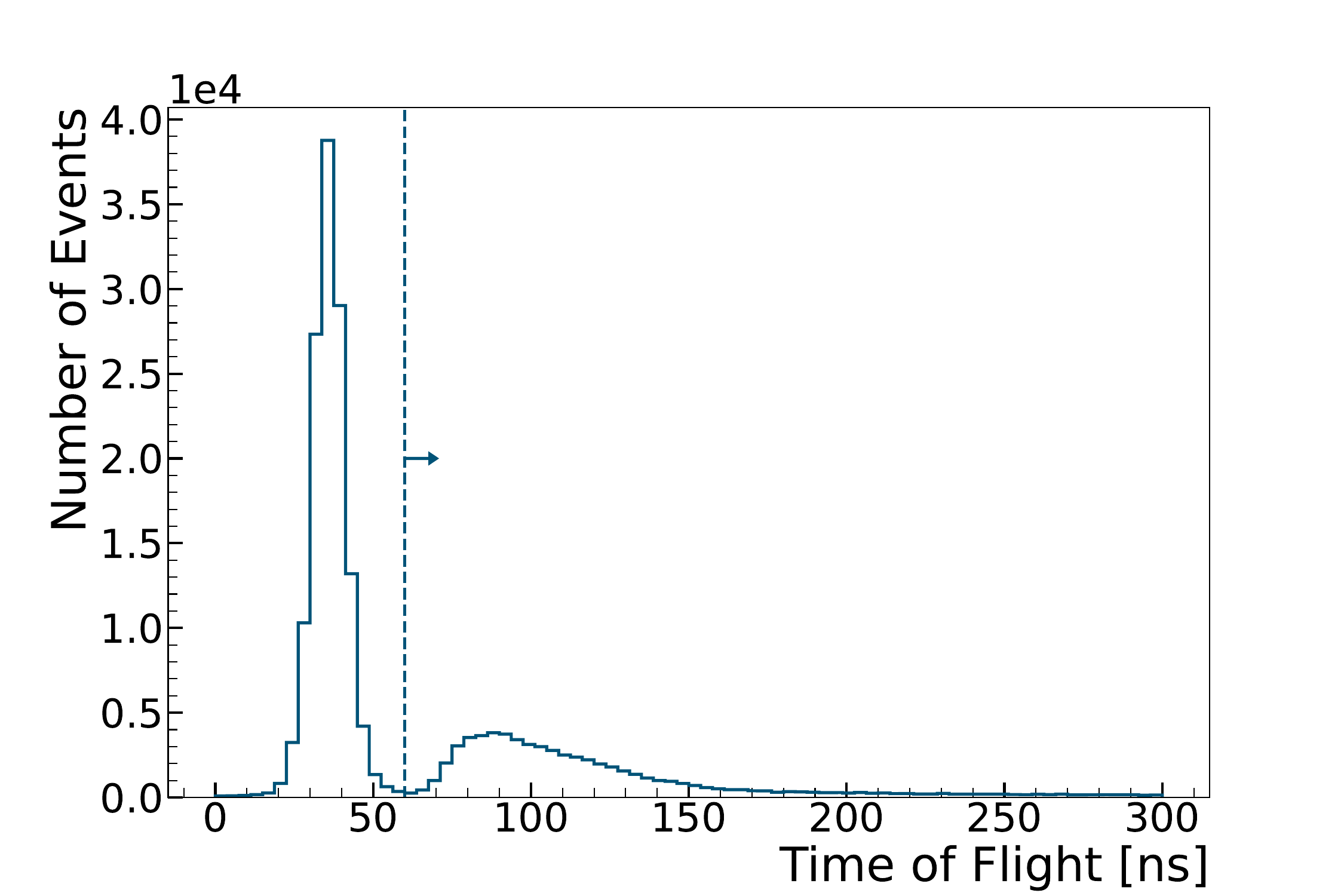}
    \includegraphics[width=0.495\textwidth]{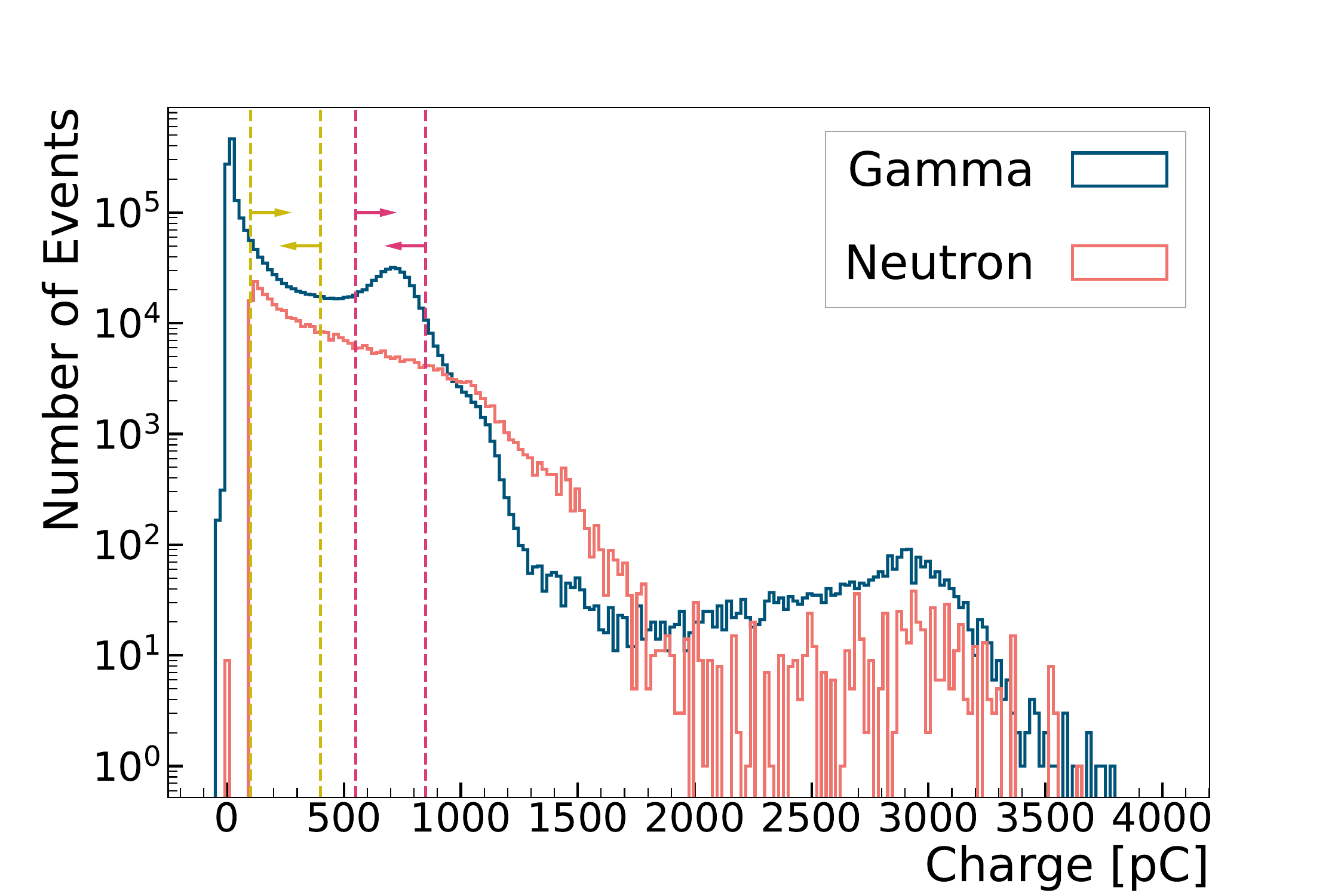}
    \caption{Left: The time-of-flight spectrum for the dataset collected using an Am-BE source and muon detector. The initial peak would consist of a mix of gamma rays and cosmogenic muons, whilst the second peak consists of neutrons travelling at a lower velocity due to their mass. The selection cut is shown. Right: The charge spectra for the gamma/neutron datasets, with the windows for each energy scale shown. The neutron dataset has already had the time-of-flight selection applied.}
    \label{fig:PSDEventSelec}
\end{figure}

To select a neutron sample, the time-of-flight was required to be > 60~ns (shown in Fig.~\ref{fig:PSDEventSelec}), which is determined by the difference between waveform trigger time in the prototype detector and in the muon detector after accounting for PMT and cabling delays. No selections were performed exclusively to the gamma ray sample. 

The remaining selection criteria are applied to both samples in the same way. Two windows are applied to the charge, which allow us to split our samples into an MeV-scale sub-sample and a keV-scale sub-sample. The first window surrounds the main 1.17/1.33~MeV $^{60}$Co gamma ray peak, whilst the second window is in the low charge region of the spectrum, both windows are applied to both samples. We can be confident there are populations of both particles in this second window. The Am-Be neutron energy spectra extends low~\cite{AmBeSpec}, and there should be a high presence of background gammas in our $^{60}$Co dataset as the detector is unshielded and above ground. Other criteria include requirements on the pulse start for each event, to remove events where the pulse occurs too close to either the start or end of the waveform (and may not have their charge accurately measured), and we remove noisy waveforms based on the standard deviation of the baseline in the first 80~ns of the waveform. Summarising, the criteria are: a MeV-scale charge window of $550<Q<850$~pC, a keV-scale charge window of $100<Q<400$~pC, shown in Fig.~\ref{fig:PSDEventSelec}; requiring pulse's start between $335<t_{\rm start}<750$~ns; and that the baseline standard deviation be $< 7$ ADC.
Pulse shape variables are calculated for the events in each sample that pass the selection criteria and are calculated for events with pulse charges within the charge window detailed above. 

\subsection{Pulse Shape Variables}
\label{sec:psd}

Pulse shape variables are used to discriminate between particle or event type by exploiting the differences in the shape of light or charge pulses from a detector, which arise due to separate physical processes that give rise to the pulse. Here we demonstrate discrimination between pulses originating from gamma rays, and from neutrons. Gamma rays, upon interacting with the scintillator, generate pulses with shorter tails because of the faster light emission process (direct interaction with electrons in the organic scintillator, causing light emission by excitation and de-excitation). Neutrons, however, instigate pulses that have longer tails, due to the slower light emission process (driven initially by a nuclear scatter, that induces the same electronic excitation and de-excitation process). We exploit these differences by using variables such as the ones defined below. 

For our study, to discriminate between neutrons and gamma rays, five definitions of pulse shape variables (and nine variables in total) were used. These variables are: the amplitude weighted mean time of the pulse, which is defined as
    \begin{equation}
        \mu_t = \frac{\Sigma_i A_i t_i}{\Sigma_i A_i},
    \end{equation}
where $A_i$ is the amplitude if the $i$th sample, and $t_i$ is the sample's corresponding time in the waveform; a ratio of two charges in the tail of the pulse: 
    \begin{equation}
        Q_{\rm ratio} = \frac{Q_{\rm prompt}}{Q_{\rm delayed}},
    \end{equation}
where $Q_{\rm prompt}$ is the charge within a window ranging from the peak location of the pulse to 34~ns after said peak, and $Q_{\rm delayed}$ ranges from the peak location to 486~ns after the peak (these windows are chosen from inspection of the average waveforms); the charge accumulated pulse~\cite{ANAISCAP}, or CAP$_x$, variable, which is defined as
    \begin{equation}
        \text{CAP}_x = \frac{\sum_{t=t_{0} \text{ ns}}^{t=x \text{ ns}} A(t)}{\sum_{t=t_{0} \text{ ns}}^{t=t_{max} \text{ ns}} A(t)},
    \end{equation}
where $A(t)$ is the amplitude of the waveform at a given time $t$, $t_0$ is the location of the pulse's peak, and $t_{max}$ is the end of the pulse window; the skew of the pulse; and the kurtosis of the pulse.

In addition to the other four variables, five CAP$_x$ variables are defined: CAP$_{25}$, CAP$_{50}$, CAP$_{100}$, CAP$_{150}$, and CAP$_{200}$, which evenly fill out the full charge window.

\subsubsection{Performance}
\begin{figure}[htb]
    \centering
    \includegraphics[width=0.7\textwidth]{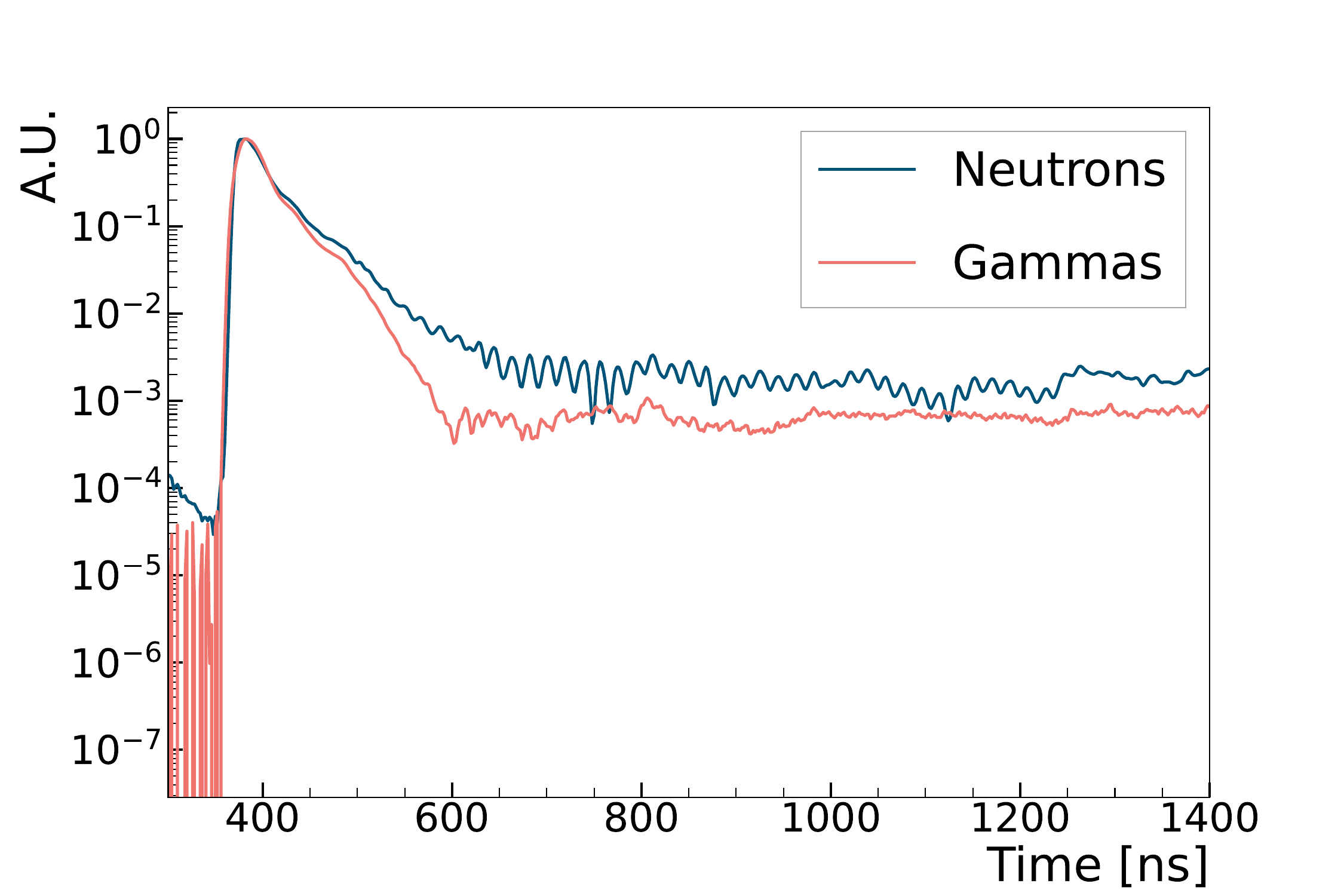}
    \caption{The peak-normalised average waveforms for the gamma ray and neutron samples, for the MeV-scale. The keV-scale version is not substantially different, with only a small change in the amount of tail charge noticeable.}
    \label{fig:AveWaveforms}
\end{figure}

By inspection of the average waveforms for both gamma rays and neutrons, which are shown in Fig.~\ref{fig:AveWaveforms}, appreciable discrimination power can be expected given the faster decay time of the gamma ray pulses, with more charge in the tail of the neutron pulses between 500~ns to 600~ns. However, the relative performance of each variable needs to be compared.

\begin{figure}[htb]
    \centering
    \includegraphics[width=0.495\textwidth]{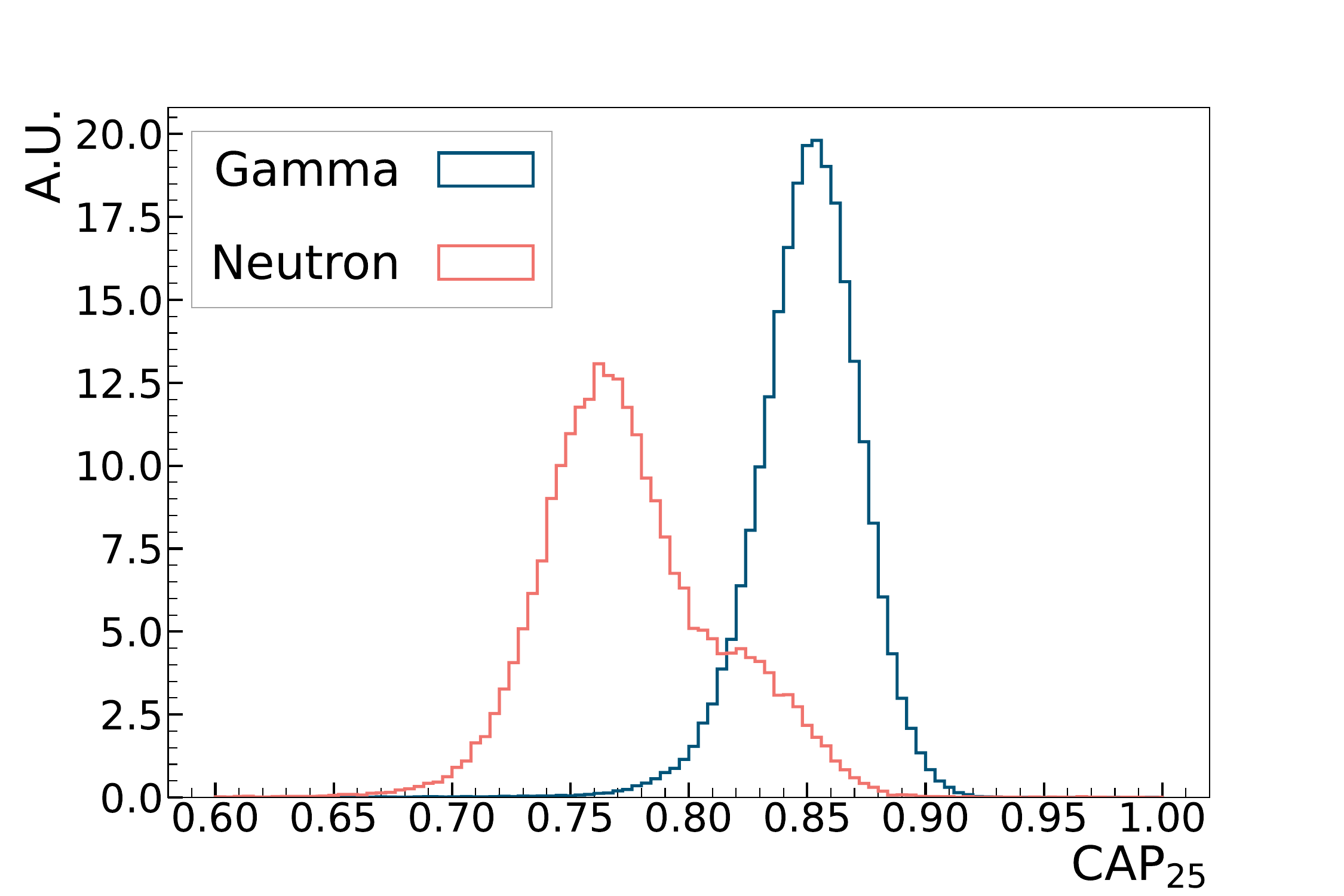}
    \includegraphics[width=0.495\textwidth]{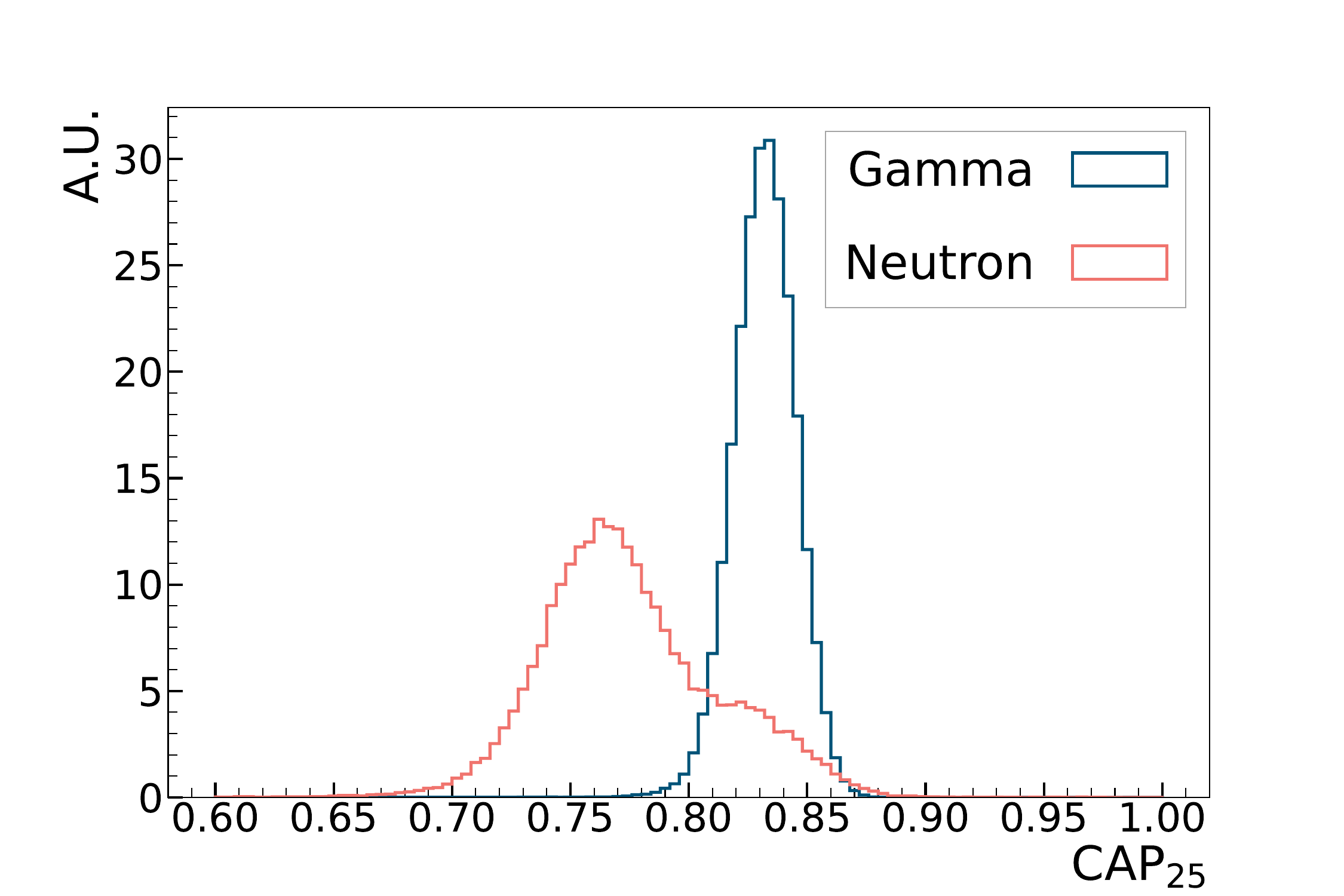}
    \caption{Left: The CAP$_{25}$ variable for the keV-scale. Right: The CAP$_{25}$ variable for the MeV-scale. Note that these histograms have been normalised. This ranking is based on the gain of a boosted decision tree. The bi-modal nature of these distributions seems to come from different timings of reflections caused by the Lumirror, where more prompt reflections inflate the amount of charge at the start of the tail, and delayed reflections have the opposite effect.}
    \label{fig:Top3Vars}
\end{figure}

We assess the performance of the variables relative to each other by feeding each into a boosted decision tree (BDT), trained to separate between gamma rays and neutrons. The BDT can rank variables based on their overall contribution to the discrimination power of classifier, using a property called gain. This gain ranking is used. We find in both the MeV and keV cases, the variables showing the highest degree of separation are CAP$_{50}$, CAP$_{25}$, and charge ratio variables (in order of degree of separation). Histograms of CAP$_{25}$, for both the keV-scale and MeV-scale are shown in Fig.~\ref{fig:Top3Vars} for comparison.

In comparison, the keV-scale variables perform worse because of a relatively higher spread in gamma ray/neutron distributions for each variable, thus reducing separation power. This is expected, as lower energies will yield lower counts of PEs in each pulse. Discrimination between gamma rays and neutrons is still viable, however. Similarly, just as the performance of the individual variables dropped, the performance of the keV-scale trained model was worse in comparison to the MeV-scale trained model.

\subsection{Frequency Domain Variables in a Low Amplitude Dataset}

It is likely that a non-negligible contribution to the rate of the veto detector may come from low amplitude electronic, or radio-frequency (RF), noise. This is of most concern in low PE datasets. We develop a variable definition to be applied in the frequency domain, and apply it to both a dark rate dataset and a SPE dataset from one of the PMTs tested in this paper. Given dark rate events and SPE events are physically indistinguishable, any separation a variable could provide should be discriminating based on other noise events in the dark rate dataset. Likewise for the pulse shape variables, we determine its efficacy in comparison to the variables tested in Sec.~\ref{sec:psd} by utilising a BDT.

We define the frequency domain variable in terms of a Fourier transform applied to a waveform, which has been smoothed with a moving average of window size $ma$, giving a spectrum $f_i^{ma}$. The smoothed spectrum is divided into bins, and for bin $x$ the mean of the spectrum is calculated, yielding $\left\langle f^{ma}_{i,x} \right\rangle$. This mean value is compared to the average of the same mean value in bin $x$ for $N$ events by means of a ratio (i.e. the dataset average of that bin value), giving us the variable $F^{ma,X}_{i,x}$, which we can define mathematically as

\begin{equation}
    F^{ma,X}_{i,x} = \frac{\left\langle f^{ma}_{i,x} \right\rangle }{\sum_i \left\langle f^{ma}_{i,x} \right\rangle/N},
\end{equation}

where $X$ corresponds to the total number of bins used to split the frequency spectrum. Qualitatively, this variable represents how much the mean amplitude of a frequency band in any given event deviates from the dataset average. This means the variable is able to pick up on noise events which give rise to narrow band peaks in a waveforms Fourier spectrum.

\begin{figure}[htb]
    \centering
    \includegraphics[width=0.6\textwidth]{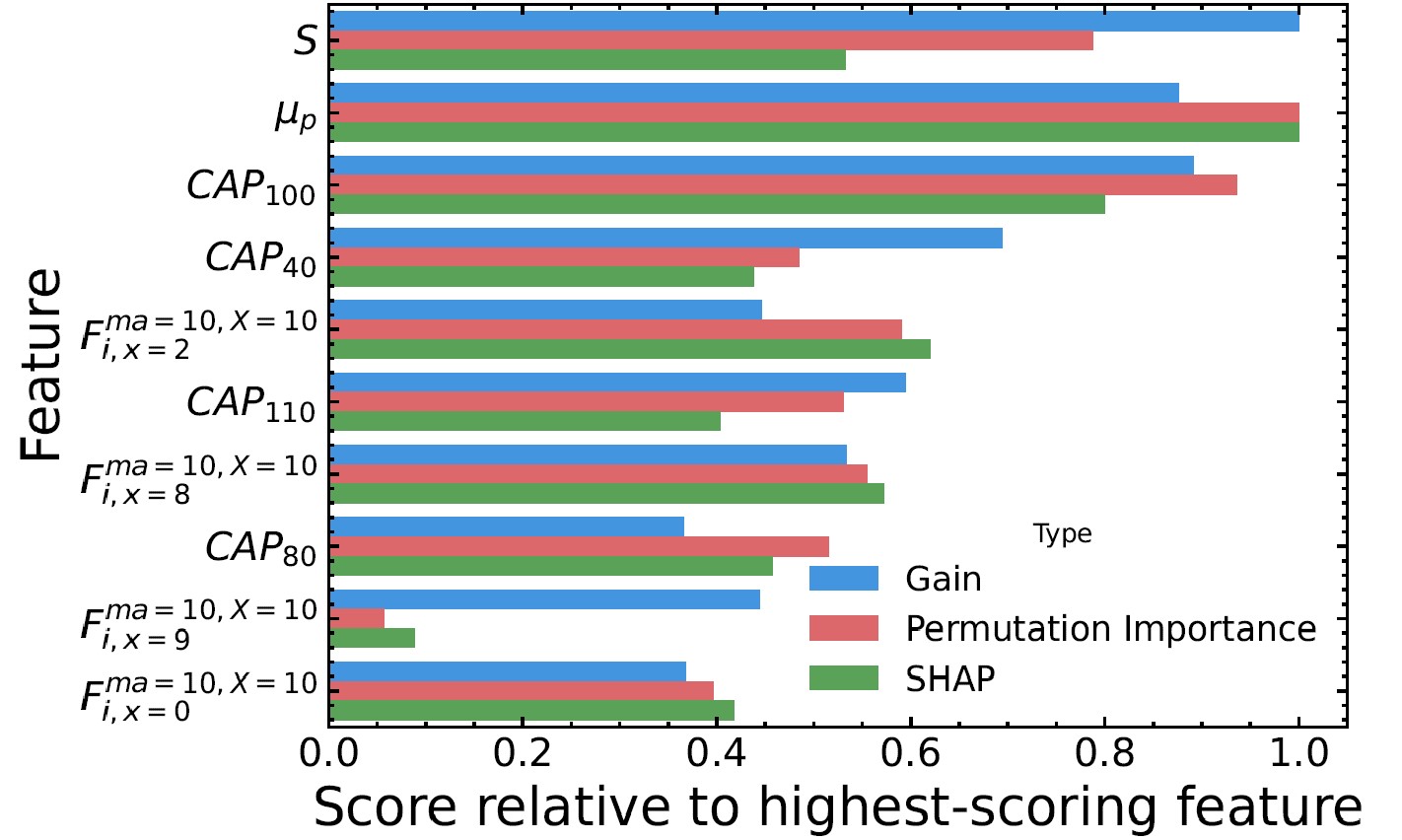}
    \caption{Feature scores for the top 10 highest scoring features in the BDT relative to the highest-scoring feature using three types of scoring method: gain, permutation importance, and SHAP.}
    \label{fig:freqvarranking}
\end{figure}

In addition to a number of frequency-based variables following the definition above, we also utilise $\rm CAP_x$ variables (using the same ones in Sec.~\ref{sec:psd}, as well as $x=40, 80, 110$), a mean time variable $\mu_p$ and the skew $S$ of the pulse within a waveform. The ranking of the top 10 variables, using three different scores, is shown in Fig.~\ref{fig:freqvarranking}. The BDT gain, permutation importance and Shapley additive explanation (SHAP) is also used. The permutation importance is model-agnostic, and scores variables according to how the performance of a model drops if that variable is shuffled within the model/tree. The SHAP variable~\cite{SHAP} quantifies the amount and direction (in terms of binary classification) of the contribution each variable provides, for each individual event. 

We can see from Fig.~\ref{fig:freqvarranking} that versions of $F^{ma,X}_{i,x}$, if defined appropriately, provide discrimination power on par with charge-based PSD variables.

\subsection{Applicability to the SABRE South Liquid Scintillator Veto}

The frequency domain variables were tested on a dark rate and SPE dataset, which is applicable to the veto detector, as low-amplitude signals are expected to make up a large amount of the potential rate. Thus, the variable defined in this paper could be used as a method to remove electronic noise from the data, either in the later analysis stages or in the initial data compression stage. 

From simulation, there is a 0.75~PE/keV position averaged detection probability for detection by any given PMT in the veto system. Using this, we can say a $\sim$100--200~PE signal would correspond to a 133--266~keV energy deposit. Our analysis of the pulse shape variables, using 1 PMT in the prototype detector, showed that even in the keV-scale window of 100--400~pC, which is 112--449~PE, neutron and gamma ray discrimination was still possible via a single variable. In this range, the best variable had a $\sim$80\% efficiency at $\sim$10\% false positive rate, whilst the BDT resulted in a $\sim$90\% efficiency at the same false positive rate.  We can infer then that pulse shape variables will be effective in the veto down to a threshold of 100s of keV. This could be be limited by the geometry of the detector, as not all PEs will be detected by a single PMT. With the application of machine learning techniques like a BDT model similar to the one utilised in this paper, or more complex neural network-based approach, this lost discrimination power could be recovered. The results from the machine learning study presented here would suggest a decrease in discrimination power with a decrease in PE statistics. However, the baseline strong performance within the prototype detector may indicate that the drop in performance when applied to the liquid scintillator veto might not be drastic.

\section{Conclusion}

The SABRE South experiment, located in the Stawell Underground Physics Laboratory (SUPL), aims to provide a Southern Hemisphere test of the DAMA/LIBRA dark matter annual modulation. A unique aspect of the SABRE South detector is the liquid scintillator veto detector, which serves the main purpose of vetoing coincident backgrounds from the NaI(Tl) targets, . We assess how efficient the veto system is at detecting energy deposits due to $^{40}$K decays in the crystal, which is the main background that will be vetoed. High efficiencies were found for a range of threshold requirements, which depend on low PE thresholds and coincidences between PMTs. For most thresholds, energy deposits above 50~keV are guaranteed to be detected, whilst the less strict thresholds can veto all deposits above 20~keV. The overall detection probability for the veto was found to be 0.75~PE/keV. To achieve this sensitivity, and apply the threshold requirements discussed, the R5912 PMTs that instrument the veto detector will need to be thoroughly pre-calibrated. We present the results and methods from a pre-calibration campaign, in which key parameters like mean SPE, gain, timing information, dark rate, and relative quantum efficiency are calibrated for up to 20 PMTs to be used in the veto. Other properties are also calibrated, which will have greater impact on future reconstruction studies using the detector. Finally, we assess the signal processing performance of an R5912 PMT within a prototype liquid scintillator detector. We analyse pulse shape discrimination performance between neutrons and gamma rays, in particular, and in two energy ranges. It is found that this technique is viable for the liquid scintillator veto, but it loses performance at lower energies. A variable designed for the removal of electronic noise is also explored, which shows promising results. The applicability of the pulse shape studies to the veto is discussed, with an approximate threshold for effective performance determined.

\section*{Acknowledgements}
The SABRE South program is supported by the Australian Government through the Australian Research Council (Grants: CE200100008, LE190100196, LE170100162, LE160100080, DP190103123, DP170101675, LP150100705). This research was partially supported by Australian Government Research Training Program Scholarships, and Melbourne Research Scholarships. This research was supported by The University of Melbourne's Research Computing Services and the Petascale Campus Initiative.


\bibliographystyle{jhep}
\bibliography{PMT.bib}

\end{document}